\documentclass[a4paper,11pt]{article}
\usepackage{jheppub} 
\usepackage{lineno}
\usepackage{xspace}
\usepackage{mathrsfs}

\newcommand{\zbl}[1]{\textcolor{black}{#1}}
\newcommand{\zblt}[1]{\textcolor{black}{#1}}

\newcommand{\UUniversal}{U(1)$_\mathrm{Universal}$\xspace}
\newcommand{\UBL}{U(1)$_\mathrm{B-L}$\xspace}
\newcommand{\ULTau}{U(1)$_\mathrm{L_\mu-L_\tau}$\xspace}

\arxivnumber{2403.05819} 

\title{Dark matter, CE$\nu$NS and neutrino new physics scrutinized by a statistical method in Xenon-based experiments}

\author{Jian Tang}
\author{and Bing-Long Zhang}
\affiliation{School of Physics, Sun Yat-sen University,\\
Guangzhou 510275, China}

\emailAdd{tangjian5@mail.sysu.edu.cn}

\abstract{
Dark matter direct detection experiments are approaching the neutrino floor, with a significant probability of measuring coherent elastic neutrino-nucleus scattering (CE$\nu$NS) and exploring potential neutrino-related new physics ($\nu$NP). In the present study, the simultaneous presence of dark matter and $\nu$NP is emphatically investigated, revealing a response similar to Standard Model neutrino backgrounds in Xenon-based dark matter experiments. 
Through analyses of three U(1) extension models, it is determined that dark matter signals can be differentiated from an excess or a depletion of neutrino contributions from $\nu$NP by applying a statistically defined distinction method to nuclear and electronic recoil spectra. 
Additionally, an investigation is conducted into how $\nu$NP affects the exclusion limits for spin-independent dark matter-nucleon interactions. The present findings could facilitate the identification of new physics in future dark matter experiments.
}

\begin{document}
\maketitle
\flushbottom

\section{Introduction}
\label{sec:intro}

The search for dark matter (DM) remains one of the most compelling endeavors in modern physics, promising profound insights into the fundamental nature of the universe~\cite{Bertone:2004pz,Roszkowski:2017nbc}. This explains the extensive efforts by physicists to construct numerous DM direct detection experiments on Earth, despite the absence of a definitive and positive DM recoil signal. 
There is a sustained push to improve low-background and low-threshold particle detection technologies with the goal of achieving the first direct detection of DM in an underground laboratory. Next-generation DM direct detection experiments, including those utilizing multi-ton noble liquid detectors, are nearing the so-called neutrino floor, which is caused by Coherent Elastic Neutrino-Nucleus Scattering (CE$\nu$NS) in the Standard Model (SM).
Recently, the XENON and PandaX collaborations have demonstrated their capability to detect CE$\nu$NS events originating from solar neutrinos~\cite{XENON:2020gfr, PandaX:2022aac}. As such, there has been a surge of interest in exploring the implications of neutrino backgrounds in DM detectors, where the concept of the neutrino floor has been proposed~\cite{Monroe:2007xp, Strigari:2009bq, Billard:2013qya, Ruppin:2014bra,Dent:2016iht, Wyenberg:2018eyv, OHare:2020lva, OHare:2021utq, Akerib:2022ort, Tang:2023xub, Herrera:2023xun}.
The neutrino floor quantifies the statistical obstruction on DM searches from neutrino backgrounds. Specifically, for Xenon-based experiments, the neutrino floor for DM masses below approximately 10~GeV is primarily determined by solar $^8$B neutrinos, while for higher DM masses, it is dominated by atmospheric neutrinos and supernova relic neutrinos (or diffuse supernova neutrino backgrounds).
Due to the higher fluxes of solar neutrinos, DM searches at lower mass ranges are more vulnerable to these backgrounds, resulting in a significantly elevated neutrino floor. Conversely, the ability to detect solar neutrinos in DM detectors also presents an opportunity to investigate neutrino-related new physics ($\nu$NP). Hence, next-generation DM direct detection experiments serve as effective astrophysical neutrino observatories~\cite{DARWIN:2016hyl,PandaX:2024oxq}.

The recent observation of CE$\nu$NS represents a major milestone in the precise measurement of SM processes~\cite{COHERENT:2017ipa, COHERENT:2020iec, COHERENT:2021xmm}. CE$\nu$NS has rapidly become a crucial tool for investigating new physics related to neutrinos~\cite{Barranco:2005yy, Farzan:2018gtr, Brdar:2018qqj, AristizabalSierra:2018eqm, CONNIE:2019xid, Cadeddu:2020nbr, delaVega:2021wpx, CONUS:2021dwh, AtzoriCorona:2022moj, Abdullah:2022zue, DeRomeri:2022twg, Breso-Pla:2023tnz, Chatterjee:2024vkd}, particularly in constraining non-standard neutrino interactions mediated by light particles. The presence of light mediators leads to more significant effects at lower energies, making DM experiments with their lower energy thresholds more sensitive to neutrino new interactions mediated by light particles. 
Notably, the low energy excesses observed by several DM experiments~\cite{Fuss:2022fxe} might indicate the existence of DM and $\nu$NP.
With large-scale DM detectors either operational or in development, detecting CE$\nu$NS events and searching for $\nu$NP within the non-standard interaction~\cite{Dutta:2015vwa, DARWIN:2016hyl, Bertuzzo:2017tuf, Aalbers:2022dzr, Amaral:2023tbs,Schwemberger:2023hee} and light mediator~\cite{Amaral:2020tga,Amaral:2021rzw,Majumdar:2021vdw,Schwemberger:2022fjl,Li:2022jfl, Demirci:2023tui,DeRomeri:2024dbv} regimes in DM experiments are becoming increasingly feasible. It will be particularly interesting to examine how $\nu$NP influences DM searches.

Recent studies have demonstrated that the neutrino floor or the discovery limit of DM could be enhanced or suppressed by various \zbl{$\nu$NP} scenarios, such as non-standard interactions and light mediator models~\cite{AristizabalSierra:2017joc,Gelmini:2018ogy,Gonzalez-Garcia:2018dep,Boehm:2018sux,Chao:2019pyh,Sadhukhan:2020etu,AristizabalSierra:2021kht}. 
Yet, there is still a gap in the comprehensive analysis of the interplay among events from DM, CE$\nu$NS, and $\nu$NP in the DM detector~\footnote{In this article, we refer to CE$\nu$NS as the \zbl{SM} predicted coherent elastic neutrino-nucleus scattering only.}.
Notably, $\nu$NP can either deplete or enhance neutrino event rates. In the case of depletion, DM signals might compensate for a reduction in neutrino events caused by $\nu$NP, analogous to scenarios where only CE$\nu$NS backgrounds are present. Therefore, a lack of excess events in the detector does not necessarily indicate an absence of DM signals, which warrants critical investigation.
Conversely, when $\nu$NP increases neutrino event rates, these additional events may resemble DM signals. Thus, it is crucial to distinguish between DM signals and CE$\nu$NS backgrounds and determine whether any observed excess is due to DM or $\nu$NP. 
Moreover, calculating the neutrino floor is challenging because the maximal likelihood ratio method is computationally intensive when considering uncertainties from neutrino fluxes, astrophysical observations, and detector response models. 
To date, there has been limited exploration of scenarios where reconstructed spectra include contributions from DM, CE$\nu$NS and $\nu$NP in an experiment. The aim of the present study is to advance this area by focusing on DM as the protagonist and examining the impact of $\nu$NP on DM searches.

In this paper,  the concept of  ``distinction limit" is proposed to facilitate two key discussions: i) distinguishing the simultaneous presence of DM and $\nu$NP from the SM interaction only, denoted by $\rm DM+\nu NP~v.s.~SM$, to assert the discovery of DM and $\nu$NP; ii) distinguishing between the presence of DM and $\nu$NP, denoted by $\rm DM~v.s.~\nu NP$, to claim the discovery of DM. For illustrative purposes, $\nu$NP is specifically demonstrated by popular U(1) models with light mediators, such as \UUniversal, \UBL and \ULTau.
The distinction limit aligns with the discovery limit used to define the neutrino floor. It is also based on the maximal likelihood ratio, with its asymptotic behavior being easily derivable as demonstrated in our recent work~\cite{Tang:2023xub}. In this study, an efficient calculation method is established for the distinction limit, enabling more convenient analyses for the simulated data sets.
While most researches on the neutrino floor have concentrated on DM-nucleon interactions, and some have addressed DM-electron interactions~\footnote{Throughout this paper, DM-nucleon interactions refer to spin-independent DM-nucleon interactions, and DM-electron interactions represent the interaction mediated by a heavy vector mediator.}~\cite{Essig:2018tss, Wyenberg:2018eyv, Carew:2023qrj}, few studies have explored the impact of $\nu$NP on the neutrino floor in the context of DM-electron interactions. Therefore, an investigation is also conducted into how $\nu$NP influences DM searches specifically for DM-electron interactions. The relevant code for this analysis is publicly available on GitHub~\cite{ourcode}.

This article is organized as follows: In Section~\ref{sec:Method}, an overview of recoil spectra on DM-nucleus interactions, DM-electron interactions, CE$\nu$NS and its new physics $\nu$NP predicted by three U(1) models is presented. The statistic method for the distinction limit is given at Section~\ref{sec:Eistinct}. In Section~\ref{sec:application}, we apply the statistic method to the mock data to demonstrate the influence of $\nu$NP on DM searches in four aspects: a distinction of DM+$\nu$NP versus SM-only assumption; a competition of DM and $\nu$NP; impacts of $\nu$NP on DM exclusion limits; model-dependent interpretations.
The findings of this work are finally summarized in Section~\ref{sec:conclusion}.

\section{Overview of recoil spectra}
\label{sec:Method}
In order to present mock data for Xenon-based DM experiments, we first review the measurable spectra for DM-nucleus and DM-electron recoils, CE$\nu$NS scattering in SM and its new physics predicted by three popular U(1) models.
\zbl{The elastic neutrino-electron scattering process is not included in our work, since its contributions to low energy events can be safely neglected~\cite{Essig:2011nj}.}

\subsection{DM-nucleus recoil}
In this paper, we concentrate solely on spin-independent (SI) DM-nucleon interactions, which has garnered significant attention in the physics community. The differential DM-nucleus scattering rate is expressed as:
\begin{equation}
	\frac{dR}{dE_r}=\frac{\rho_\chi A^2}{2 m_\chi \mu_N^2}\sigma_{\chi-p}^0 F^2(E_r) \int_{v_{min}(E_r)}^{v_{esc}}{\frac{f(v,v_0)}{v}d^3v}\,.
\end{equation}
where $E_r$ is the nuclear recoil (NR) energy, $\sigma_{\chi-p}^0$ is the SI DM-nucleon cross section, $m_\chi$ is the DM mass, $\mu_N$ is the reduced mass for the DM and nucleon collision, $A$ is the atomic number, $F^{2}\left(q^{2}\right)$ is the form factor of the nucleus, with the Helm form being assumed here~\cite{Helm:1956zz} and the momentum transfer $q$. With regards to other parameters, we adopt the standard halo model (SHM). The local DM density is $\rho_\chi = 0.3~\mathrm{GeV / cm}^3$, the circular velocity of the Local Standard of Rest is $v_0=220~\mathrm{km/s}$ , the escape velocity in the Milky Way is $v_{esc}=544~\mathrm{km/s}$, and $f(v,v_0)$ stands for the DM velocity distribution~\cite{Drukier:1986tm,Evans:2018bqy}. $v_{min}(E_r)$ is the minimal velocity required for the recoil between DM and the target nucleus. 
Given that the data analysis for $^8$B neutrino searches in DM direct detection experiments have lowered the energy threshold to $\lesssim 1~\mathrm{keV}$~\cite{XENON:2020gfr, PandaX:2022aac}, we set the energy threshold to $E_r=1~\mathrm{keV}$ and disregard the NR detection efficiency unless it is necessary such as the discussion in Part~\ref{sec:exclusion}. 

\subsection{DM-electron recoil}
Similar to the references~\cite{Essig:2011nj,Essig:2015cda}, we express the DM electronic recoil (ER) as follows:
\begin{equation}
\frac{d\langle\sigma_{ion}^{nl}\rangle}{d\ln E_{er}}=\frac{\overline\sigma_e}{8\mu_{\chi e}^2}\int qdq |f_{ion}^{nl}(k',q)|^2|F_{\rm{DM}}(q)|^2\eta(v_{min})\,,
\end{equation}
where $E_{er}$ is the electron recoil energy, $\mu_{\chi e}$ is the reduced mass for the DM and electron collision, $\eta(v_{min})$ is the inverse mean speed for a given velocity distribution, and $v_{min}$ is the minimal velocity for the momentum transfer $q$ and the outgoing electron momentum $k'=\sqrt{2m_e E_{er}}$ . $|f_{ion}^{nl}(k',q)|^2$ stands for $(n,l)$ shell ionization form factor which can be calculated by wave functions in Schrodinger equations. $\overline\sigma_e$ is the cross section given a momentum transfer $q=\alpha m_e$, where $q$-dependence is described by the DM form factor in $F_{\rm{DM}}(q)$. Here we only consider the heavy vector mediator case $F_{\rm{DM}}(q)=1$ for the sake of simplicity.

The differential ionization rate is expressed as:
\begin{equation}
\frac{dR_{ion}}{d\ln E_{er}}=\frac{1}{m_N}\frac{\rho_{\chi}}{m_{\chi}}\sum_{nl}\frac{d\langle \sigma_{ion}^{nl}v\rangle}{d\ln E_{er}}\,,
\end{equation}
where $m_N$ is the mass of the target nucleus. In this paper, we follow the procedure outlined in Refs.~\cite{Essig:2012yx,Essig:2017kqs} to convert the electron recoil energy $E_{er}$ into the observable electron yield $n_e$. 
This conversion method is also applied to derive neutrino event rates in the following section, while ionization yield function is needed to convert the nuclear recoil energy into $n_e$.
To mitigate uncertainties associated with the ionization yield function and the model of emitted electrons for Xenon, we set the threshold to $n_e=3$, consistent with experimental findings~\cite{XENON:2019gfn,PandaX-II:2021nsg,PandaX:2022xqx}.

\subsection{Coherent elastic neutrino-nucleus scattering}
\label{sec:cevns}
In \zbl{SM}, the differential cross section of the CE$\nu$NS process is given by:
\begin{equation} \label{eqn:v-crosssection}
\frac{d \sigma_{\nu_l-N}\left(E_{\nu}, E_{r}\right)}{d E_{r}}=\frac{G_{F}^{2} m_N}{\pi}  (Q_{\ell, \mathrm{SM}}^{V})^2 F^{2}\left(E_r\right) \left(1-\frac{m_N E_{r}}{2 E_{\nu}^{2}}\right) \,,
\end{equation}
where $E_r$ is the nuclear recoil energy, $E_\nu$ is the incoming neutrino's energy, $G_F$ is the Fermi constant. The weak charge can be written as:
\begin{equation} 
Q_{\ell, \mathrm{SM}}^{V} = (2Z+N) g_V^u + (2N+Z) g_V^d=-(N-Z(1-4 \sin^2\theta_W))/2 \,.
\end{equation}
where $g_V^u=\frac{1}{2}-\frac{4}{3}\sin^2\theta_W$,  $g_V^d=-\frac{1}{2}+\frac{2}{3}\sin^2\theta_W$, and $\theta_W$ is the weak-mixing angle.

The neutrino spectrum in the DM detector is provided by:
\begin{equation}\label{eqn:v-spectrum}
	\frac{d R}{d E_{r}}=\frac{1}{m_N} \sum_{i,\ell} \int_{E_{\nu}^{\min}(E_r)} \frac{d \Phi_i}{d E_{\nu}}P_{e\ell} 
 \frac{d \sigma_{\nu_l-N}}{d E_{r}}  d E_{\nu}
	\,,
\end{equation}
where $E_{\nu}^{\min}(E_r)$ is the minimal kinetic energy generated by neutrino recoils, $\frac{d \Phi_i}{d E_{\nu}}$ represents the differential neutrino flux from the known source, $P_{e\ell}$ stands for the neutrino survival probability from the Sun to the Earth. 
Our focus in this paper is primarily on solar neutrinos, which contribute the majority of events in the DM detector. In the flavor-independent case, neutrinos of different flavors elicit the same response in the DM detector, and $\nu$NP does not affect the neutrino propagation. Thus, the survival probability for different-flavor neutrinos can be out of consideration. 
However, in the flavor-dependent case, both neutrino propagation and detection are influenced by $\nu$NP, so that a careful treatment is required.
Our study specifically considers the $^8$B and hep neutrino sources, as they dominate in the region of interest.
Analogous to the conversion of electron recoil energy $E_{er}$ into observable electron yield $n_e$ in the context of DM-electron interactions scenario, we adopt a similar approach to convert the nuclear recoil energy $E_{r}$ into $n_e$. The ionization yield function is extracted from Fig.~4 in Ref.~\cite{Schwemberger:2022fjl}. To mitigate uncertainties associated with the yield function in low energy, we set the threshold $n_e=3$ and adopt the fiducial ionization yield function.

\subsection{Neutrino new physics in CE$\nu$NS}

The CE$\nu$NS predictions can be modified by neutrino new physics models which offer additional neutral-current interactions. Among them, the simplest extensions are U(1) models. 
Though the main purpose in the current study is to discriminate the potential excess or depletion caused by neutrino new physics, it would be more convincing to show the concrete model predictions in the mock data. Therefore, we focus on three popular U(1) models: \UUniversal, \UBL and \ULTau. 
For the first two cases, the Lagrangian describing the vector boson $Z^\prime$ is given by:
\begin{equation}
\mathcal{L}_{Z^{\prime}}^{V}=-g_{Z^{\prime}} Z_{\mu}^{\prime}\left[\sum_{\ell=e, \mu, \tau} Q^\prime_{\ell} \overline{\nu_{\ell L}} \gamma^{\mu} \nu_{\ell L}+\sum_{q=u, d} Q^\prime_{q} \bar{q} \gamma^{\mu} q\right]\,,
\end{equation}
where $g_{Z^{\prime}}$ is the new vector coupling constant, \zbl{$Q^\prime_{\ell}$ and $Q^\prime_{q}$ are the U(1) charge for the lepton and the quark, respectively.}
To obtain the modified spectrum, one can simply substitute the weak charge in~Eqn.~(\ref{eqn:v-crosssection}) with:
\begin{equation}
Q_{\ell, \mathrm{SM}+\mathrm{V}}^{V}=Q_{\ell, \mathrm{SM}}^{V}+\frac{g_{Z^{\prime}}^{2} Q_{\ell}^{\prime}}{\sqrt{2} G_{F}\left(|\vec{q}|^{2}+m_{Z^{\prime}}^{2}\right)}\left[\left(2 Q_{u}^{\prime}+Q_{d}^{\prime}\right) Z +\left(Q_{u}^{\prime}+2 Q_{d}^{\prime}\right) N\right]\,,
\end{equation}
where $m_{Z^{\prime}}$ is $Z^\prime$ mass, and $|\vec{q}|=\sqrt{2 m_N E_r}$ is the momentum transfer. $N$ and $Z$ are the number of neutrons and protons in the target nuclei, respectively. Please see Ref.~\cite{AtzoriCorona:2022moj} for more details.

\zbl{In \UUniversal model~\cite{Liao:2017uzy,Billard:2018jnl,Denton:2018xmq,Papoulias:2019txv,Papoulias:2019xaw,Cadeddu:2020nbr,AtzoriCorona:2022moj},} the vector mediator $Z^\prime$ couples universally to all \zbl{SM} fermions, meaning $Q^\prime_{\ell}=Q^\prime_{q}=1$. This model cannot be anomaly-free without the introduction of extra fermions from the full theory. 
Note that the additional part of $Q_{\ell, \mathrm{SM}+\mathrm{V}}^{V}$ counteracts $Q_{\ell, \mathrm{SM}}^{V}$. Therefore, the modified spectrum may vanish for proper parameters, leading to a thin allowed strip within the constraints from the COHERENT data~\cite{AtzoriCorona:2022moj}. The modified neutrino event rates get depleted with the \UUniversal model.
\zbl{The \UBL model~\cite{Langacker:2008yv, Billard:2018jnl,Mohapatra:2014yla,Okada:2018ktp,Cadeddu:2020nbr,AtzoriCorona:2022moj}} is one of the most popular $Z^\prime$ models with extensive discussions. In \UBL model, three right-handed neutrinos are introduced to ensure anomaly cancellation and provide an explanation for the origin of neutrino masses. The assignments of \UBL charges differ in quarks and leptons: $Q^\prime_{q}=1/3$ and $Q^\prime_{\ell}=-1$, which leads to an enhancement of $Q_{\ell, \mathrm{SM}+\mathrm{V}}^{V}$.

\zbl{The \ULTau model~\cite{He:1990pn,Abdullah:2018ykz,Altmannshofer:2019zhy,Cadeddu:2020nbr,Banerjee:2021laz,AtzoriCorona:2022moj}} is anomaly-free without the need for additional non-standard particles. Although the $Z^\prime$ boson in this model does not directly couple to quarks, loop effects introduce a kinetic-mixing term in the Lagrangian, enabling its contribution to neutrino-nucleus scattering. The modified spectrum for the \ULTau model is obtained by replacing the weak charge with:
\begin{equation}
Q_{\ell ,\mathrm{SM} +\mathrm{U}( 1)_{L_{\mu } -L_{\tau }}}^{V} =Q_{\ell ,\mathrm{SM}}^{V} +\frac{\sqrt{2} \alpha _{\mathrm{EM}} g_{Z^{\prime }}^{2}( \delta _{\ell \mu } \varepsilon _{\tau \mu } (|\vec{q} |) +\delta _{\ell \tau } \varepsilon _{\mu \tau } (|\vec{q} |)}{\pi G_{F}\left( |\vec{q} |^{2} +m_{Z^{\prime }}^{2}\right)} Z\,,
\end{equation}
where $\alpha _{\mathrm{EM}}$ is the electromagnetic ﬁne-structure constant, $\delta_{\ell\mu}$ is the Kronecker delta symbol and $\varepsilon_{\tau\mu}(|\vec{q}|)=-\varepsilon_{\mu\tau}(|\vec{q}|)$ is the one-loop kinetic mixing coupling approximated as $\varepsilon_{\tau\mu}\simeq\ln(m_{\tau}^2/m_{\mu}^2)/6$.
In \ULTau model, neutrinos of different flavors interact with nuclei and electrons in various behaviours, which denoted by a flavor-dependent scenario. \zbl{Therefore, both the detection in the DM experiment and the neutrino propagation from the Sun to the Earth must consider the effects of $\nu$NP. The current study focuses on neutrino scatterings. The contributions to the neutrino detection from $\nu$NP will be extensively discussed while the $\nu$NP impact on the neutrino propagation is described in Appendix~\ref{appendix3}.}

\begin{figure}[htbp]
	\centering
  \includegraphics[width=0.48\linewidth]{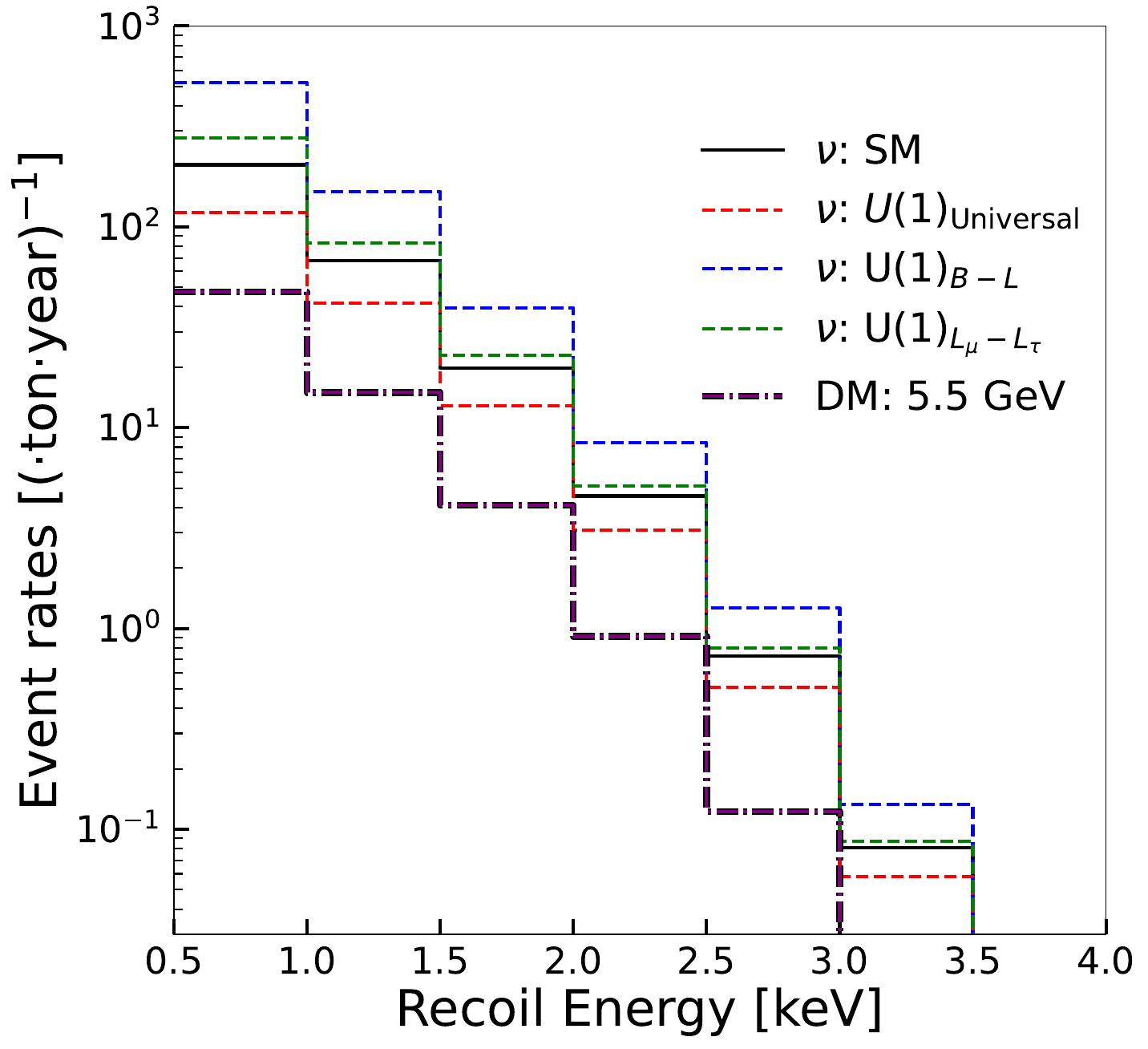}
	\includegraphics[width=0.46\linewidth]{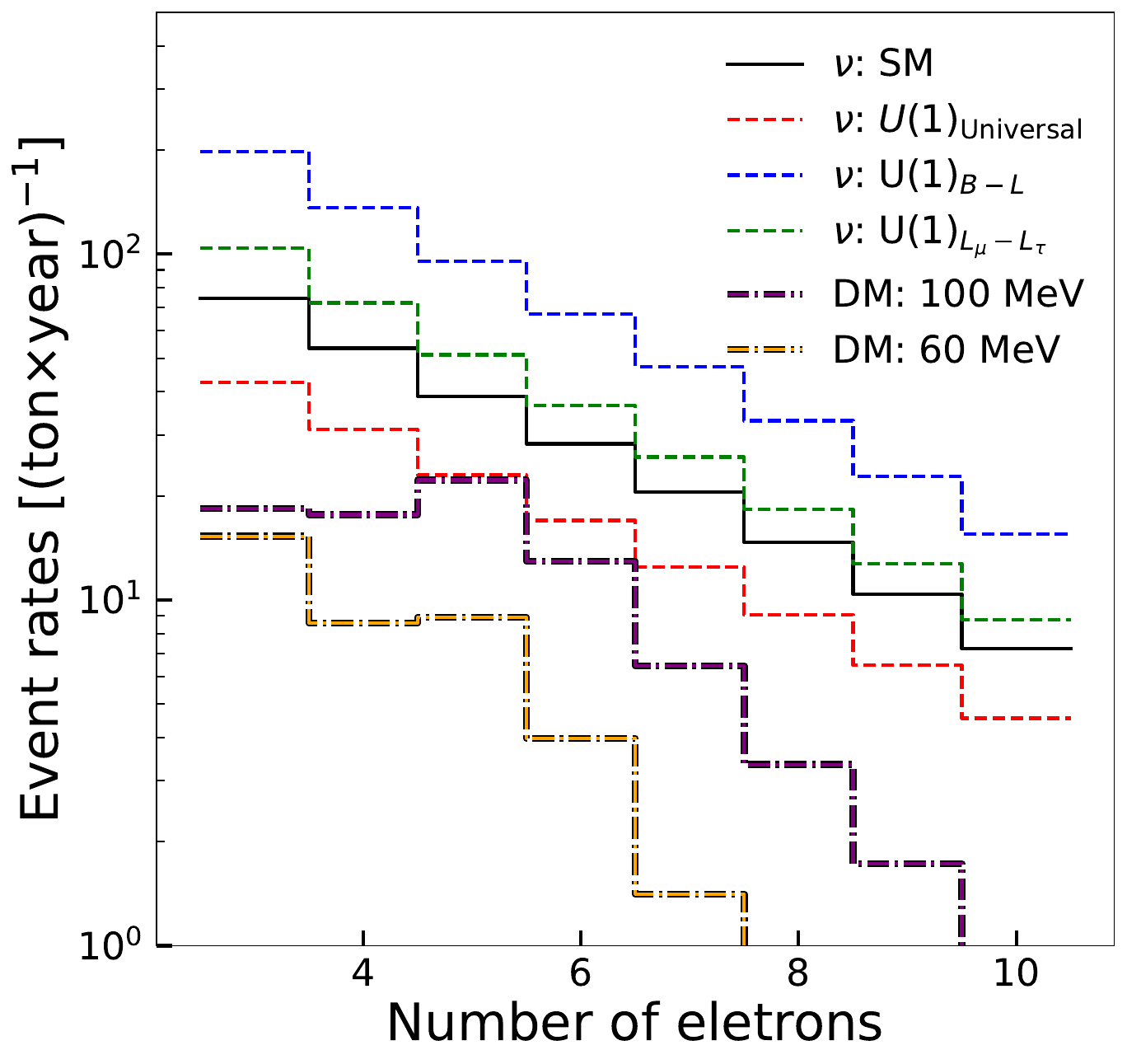}
	\caption{\label{fig:ERSpectrum} The left panel is nuclear recoil spectra in a liquid Xenon detector for neutrinos and DM. The solid line denotes neutrino spectra for SM only while dashed lines are for different U(1) extension model predictions, \zblt{with their benchmark parameters shown in Tab.~\ref{tab:benchmark}.} The DM-induced event rates are given by a dashed-dotted line\zbl{, with the DM-nucleon cross section at $1\times10^{-45}~\mathrm{cm}^2$}. The right panel is electron recoil spectra in the similar manner\zbl{, with the DM-eletron cross section at
 $2\times10^{-42}~\mathrm{cm}^2$. }}
\end{figure}

\begin{table}[htbp]
\centering
\caption{\zblt{Benchmark points for three U(1) models.}}
\label{tab:benchmark}
\begin{tabular}{|c|c|c|}
\hline
model                        & $m_{Z^\prime}~[\mathrm{MeV}]$ & $g_{Z^\prime}$      \\ \hline
U(1)$_\mathrm{Universal}$    & 30                            & $2\times10^{-5}$    \\ \hline
U(1)$_\mathrm{B-L}$          & 20                            & $4\times 10^{-5}$   \\ \hline
U(1)$_\mathrm{L_\mu-L_\tau}$ & 16                            & $8\times 10^{-4}$ \\ \hline
\end{tabular}
\end{table}

Based on the formalism, we can now compute the mock data in a typical Xenon-based DM experiment, as shown in Fig.~\ref{fig:ERSpectrum}. Let us make a comparison in neutrino and DM spectra in terms of the recoil energy transferred by DM-nucleon interactions and the number of emitted electrons for DM-electron interactions. \zblt{Given that a lighter mediator results in more substantial changes in DM experiments, we choose benchmark points within the non-excluded parameter space for three U(1) models. These benchmark points are detailed in Tab.~\ref{tab:benchmark}. Additionally, these benchmark points are displayed along with the excluded regions as illustrated in Fig.~\ref{fig:U1-constraints}.} 
One can see from Fig.~\ref{fig:ERSpectrum} that the event rates induced by neutrinos and DM are comparable, which indicates neutrinos with or without $\nu$NP interactions would significantly influence DM searches. For \UUniversal depleting neutrino event rates, the total response of DM and neutrinos would be similar to the response with an absence of both DM and $\nu$NP. On the contrary,  for \UBL and \ULTau enhancing neutrino event rates, the additional contributions from CE$\nu$NS backgrounds would be similar to DM signals. 
Therefore, it is a right question to ask how to discriminate these scenarios. \zbl{It will be interesting to see that the statistical method with a definition of the distinction limit can answer the question effectively.}

\section{Statistical methodology}
\label{sec:Eistinct}
To expound upon the statistical concept of the distinction limit, we shall first revisit the definition of discovery limit. The discovery limit was proposed to quantify the expected signiﬁcance of a DM detection taking into account astrophysical and experimental uncertainties~\cite{Billard:2011zj}, where a frequentist approach based on the proﬁle likelihood ratio test statistic was adopted.  
Formally, the discovery limit can be defined as the minimum cross section required for an experiment to have a 90\% probability or 90\% confidence level (C.L.) to detect a DM signal with a 3$\sigma$ significance~\cite{Billard:2013qya,Ruppin:2014bra}. 
Based on the discovery limit, the neutrino floor was proposed to systematically assess the discovery potential of searching for DM in the presence of the neutrino background.
In this framework, we attribute the absence of a DM signal and the presence of an existing DM signal to the null hypothesis H$_0$ and the alternative hypothesis H$_1$, respectively. According to the definition of discovery limit, the $3\sigma$ significance of signals corresponds to a p-value $p_0 = 0.0027$. The p-value $p_0$ is defined as $p_0=\int^\infty_{q_\mathrm{obs}} f(q_0|\mathrm{H}_0)dq_\mathrm{obs}$, where $f(q_0|\mathrm{H}_0)$ is the distribution of the statistic under the assumption of H$_0$ being true. The p-value is believed to follow a chi-square distribution~\cite{Wilks:1938dza,Chernoff:1954eli,Cowan:2010js}. 
The $90\%$ C.L. denotes the percentage of experiments in which the discovery is deemed significant, i.e., $\int^\infty_{q_\mathrm{obs}} f(q_0|\mathrm{H}_1)dq_\mathrm{obs}=90\%$. 

To quantify the impact of $\nu$NP on DM searches rigorously, a stringent statistical criterion is essential. \zbl{In the discussions that follow, any mention of $\nu$NP should be interpreted as implicitly encompassing the SM neutrino interaction.} In the case $\rm DM+\nu NP~v.s.~SM$, the null hypothesis H$_0$ assumes the presence of Standard Model CE$\nu$NS backgrounds only, while the alternative hypothesis H$_1$ assumes the presence of both DM and $\nu$NP induced rates. By putting together an appropriate test statistic $q$, the corresponding distributions of two cases are $f(q|\mathrm{SM})$ and $f(q|\mathrm{DM+\nu NP})$, respectively.
The influence of $\nu$NP is expected to manifest in similar spectra, potentially resulting in the failure of observing DM and $\nu$NP signals in a DM experiment, leading to the statistical statement that only CE$\nu$NS backgrounds are observed.
In the case $\rm DM~v.s.~\nu NP$, H$_0$ posits the presence of $\nu$NP without DM, whereas H$_1$ posits the presence of DM without $\nu$NP. The corresponding distributions are $f(q|\mathrm{\nu NP})$ and $f(q|\mathrm{DM})$, respectively. If we try to assert the discovery of DM when there is an excess on the spectrum, it is necessary to distinguish DM signals from $\nu$NP signals. \zbl{Please see Tab.~\ref{tab:tab1} for a concise summary of these statistical models.}

Therefore, we define the distinction limit to statistically assert the differentiation between H$_0$ and H$_1$. In our definition, the distinction limit stipulates that an experiment has at least a 90\% C.L. ($\int^\infty_{q_\mathrm{obs}} f(q|\mathrm{H}_1)dq=90\%$) to reject H$_0$ with a $1.28\sigma$ significance ($\int_{-\infty}^{q_\mathrm{obs}} f(q|\mathrm{H}_0)dq=90\%$) if H$_1$ is indeed true.
Here, we opt for a $1.28\sigma$ significance to make the distinction limit symmetric. The distinction limit remains the same if we seek to reject H$_1$ instead of H$_0$.
Note that in the case $\rm DM+\nu NP~v.s.~SM$, we assert the discovery of DM and $\nu$NP by rejecting the null hypothesis that only CE$\nu$NS backgrounds are observed. This assertion differs from the conventional definition of discovery. The difference arises from the dissimilarity in our construction of the test statistic compared to that of the definition of discovery outlined in Ref.~\cite{Cowan:2010js}. 

\begin{table}[htbp]
\centering
\caption{\zbl{Interpretations of parameters of interest and nuisance parameters.}}
\label{tab:tab2}
\resizebox{\textwidth}{!}{%
\begin{tabular}{|cc|cc|}
\hline
\multicolumn{2}{|c|}{parameters of interest}                   & \multicolumn{2}{c|}{nuisance parameters}                                           \\ \hline
\multicolumn{1}{|c|}{$\theta_1$} & strength of DM signal      & \multicolumn{1}{c|}{$\theta_3$} & normalized $^8$B neutrino flux, $\sigma_3$=0.02~\cite{Bergstrom:2016cbh}\\ \hline
\multicolumn{1}{|c|}{$\theta_2$} & strength of $\nu$NP signal & \multicolumn{1}{c|}{$\theta_4$} & normalized hep neutrino flux, $\sigma_4$=0.3~\cite{Vinyoles:2016djt}\\ \hline
\end{tabular}%
}
\end{table}

The expected event in the $i$-th bin is denoted by $n_{exp,i} = \theta_1 s_i + \theta_2 x_i + b_i$, where $s_i$ represents DM events, $x_i = \sum_j x_i^j \theta_j ~(j\geq 3)$ is the $\nu$NP induced events after subtracting the Standard Model CE$\nu$NS background, and $b_i = \sum_j b_i^j \theta_j ~(j\geq 3)$ stands for CE$\nu$NS events. Here, $b_i^j$ is the $j$-source events in the $i$-th bin, and $x_i^j$ is similar. \zbl{Note that $x_i$ measures the disparity in neutrino responses, specifically contrasting those resulting from the solely SM neutrino interaction and those arising from the combined effect of the SM interaction and $\nu$NP.} The parameters of interest $\theta_1$ and $\theta_2$ are the strengths of DM and $\nu$NP, respectively. Other nuisance parameters $\theta_j(j>2)$ represent the normalized neutrino fluxes from different sources, with distribution $\mathscr{N}_{\theta_j}(1, \sigma_j)$ having a central value at 1 and $\sigma_j$ as the standard deviation in a Gaussian distribution: $\mathscr{N}_{\theta_j}(1, \sigma_j)= \frac{1}{\sqrt{2\pi} \sigma_j} \exp{(-\frac{(\theta_j-1)^2}{2 \sigma_j^2})}$. \zbl{As shown in Tab.~\ref{tab:tab2}, we list all parameters utilized in this work, where only $^8$B and hep neutrinos are considered as discussed in Part~\ref{sec:cevns}.}
Now we introduce the profile likelihood as follows:
\begin{equation}\label{eqn:Likelihood}
L(\theta_1, \theta_2,\dots,\theta_m ) =
\prod_{i=1}^N \mathscr{P}(n_i|\theta_1 s_i + \theta_2 x_i + b_i)   \;\;
\prod_{j=2}^M \mathscr{N}_{\theta_j}(1, \sigma_j)  \;,
\end{equation}
where $\mathscr{P}$ is Poisson distribution with $\mathscr{P}(n_i|n_{exp,i})=\frac{ n_{exp,i}^{n_{i}} }{ n_{i}! } e^{- n_{exp,i} } $, $N$ and $M$ are the number of bins and parameters, respectively.
Then, the test statistic is defined as follows:
\begin{equation}\label{eqn:test-statistic}
q =- 2 \ln \lambda(\theta^0_1,\theta^0_2,\theta^1_1, \theta^1_2) = -2[\ln \lambda(\theta^0_1,\theta^0_2) -\ln \lambda(\theta^1_1, \theta^1_2)]\,.
\end{equation}
where the likelihood ratio $\lambda(\theta^0_1,\theta^0_2,\theta^1_1, \theta^1_2) = \frac{ L(\theta^0_1,\theta^0_2,
\hat{\hat{\theta}}_3,\dots,\hat{\hat{\theta}}_M) } {L(\theta^1_1, \theta^1_2,
\hat{\hat{\theta}}_3,\dots,\hat{\hat{\theta}}_M) }$ and $\lambda(\theta_1,\theta_2) = \frac{ L(\theta_1,\theta_2,
\hat{\hat{\theta}}_3,\dots,\hat{\hat{\theta}}_M) } {L(\hat{\theta}_1, \hat{\theta}_2,\hat{\theta}_3,\dots,\hat{\theta}_M)) }$ are defined for simplicity. 
\zbl{As shown in Tab.~\ref{tab:tab1}, we provide pertinent parameter selections and test statistics for two distinct cases: $\rm DM+\nu NP~v.s.~SM$ and $\rm DM~v.s.~\nu NP$.}

\begin{table}[htbp]
\centering
\caption{\zbl{Hypotheses and test statistics for two statistical models.}}
\label{tab:tab1}
\resizebox{\textwidth}{!}{%
\begin{tabular}{|c|c|c|c|}
\hline
                        & H$_0$                                & H$_1$                                 & q                        \\ \hline
$\rm DM+\nu NP~v.s.~SM$ & SM: $\theta^0_1=\theta^0_2=0$        & DM+$\nu$NP: $\theta^0_1=\theta^0_2=1$ & $-2\ln \lambda(0,0,1,1)$ \\ \hline
$\rm DM~v.s.~\nu NP$    & $\nu$NP: $\theta^0_1=0,\theta^0_2=1$ & DM+SM: $\theta^0_1=1,\theta^0_2=0$    & $-2\ln \lambda(0,1,1,0)$ \\ \hline
\end{tabular}%
}
\end{table}

Asymptotically, the log-likelihood ratio $-2\ln \lambda(\theta^0_1,\theta^0_2)$ (and $-2\ln \lambda(\theta^1_1, \theta^1_2)$) with a factor of $-2$ forms a quadratic polynomial of a normal variate~\cite{Tang:2023xub}. Thus, the test statistic $q$ defined in~Eqn.~(\ref{eqn:test-statistic}) is the sum over independent normal variate with various weights, since the quadratic term vanishes due to the subtraction of $-2\ln \lambda(\theta^0_1,\theta^0_2)$ and $-2\ln \lambda(\theta^1_1, \theta^1_2)$. For further details, please refer to Appendix \ref{appendix1}. 
Consequently, $q$ asymptotically follows a normal distribution, and its mean and standard deviation can be readily computed numerically. Note that we utilize Asimov dataset~\cite{Cowan:2010js} to calculate the mean for a better precision. To demonstrate the effectiveness of our method, we compared the distribution of $q$ derived from our method and Monte Carlo realisations as in Appendix~\ref{appendix2}. 
Given that effects caused by $\nu$NP are often marginalized in next-generation dark matter experiments, this study will pave the way for the computationally expensive analyses.

\section{Applications on mock data}
\label{sec:application}
This section aims to quantitatively present the effect of \zbl{$\nu$NP} on DM searches in the contexts of DM-nucleon and DM-electron interactions.
To measure the influence of $\nu$NP on DM searches, we need to compare some quantities in the analysis with/without the presence of $\nu$NP. Generally speaking, under the discovery exposure, an experiment has a 90\% C.L. to discover DM signals with $3\sigma$ sensitivity. However, the discovery exposure alone is insufficient to claim the discovery of DM signals when considering the presence of $\nu$NP. Therefore, we define the distinction exposure, under which an experiment has a 90\% C.L. to assert the discovery of DM and $\nu$NP signals for the case $\rm DM+\nu NP~v.s.~SM$ and to determine whether the any potential event excess originates from DM or $\nu$NP for the case $\rm DM~v.s.~\nu NP$. In this paper, we define a ratio of the distinction exposure to the discovery exposure, denoted by $n$, to quantify the impact from $\nu$NP. 
If $n\leq 1$, with the DM discovery exposure, the experiment can distinguish and exclude $\nu$NP effects. Meanwhile, if $n>1$, the experiment requires a greater exposure to make such distinctions, as their spectra are overlapped with each other. 

As we presented in the above section, a mixture of DM, CE$\nu$NS and $\nu$NP events will lead to confusion in the data analyses and physics interpretations. Demonstrated by three U(1) extensions, we illustrate different scenarios by numerical calculations and mock data analyses in various aspects in the following. 
In Part~\ref{sec:A}, the distinction regions with $n\geq1$ for $\rm DM+\nu NP~v.s.~SM$ are presented. The distinction regions indicate that an experiment needs a larger exposure to claim the existence of DM and $\nu$NP. 
In Part~\ref{sec:B}, we show the distinction regions for $\rm DM~v.s.~\nu NP$, which points to a larger exposure to determine whether a potential distortion in event spectra originates from DM or $\nu$NP. 
What is most interesting is that the distinction regions overlap with the excluded regions for DM-nucleon interactions. This fact indicates the potential and sizable effect of $\nu$NP on the current exclusion limits, which is investigated in Part~\ref{sec:exclusion}.
\zbl{Finally, to comprehensively illustrate the effects of U(1) extensions, we vary the energy thresholds and model parameters, and subsequently compare their respective impacts in Parts~\ref{sec:D} and~\ref{sec:E}.}

\subsection{$\rm DM+\nu NP~v.s.~SM$}
\label{sec:A}
\begin{figure}[htbp]
	\centering
	\includegraphics[width=0.8\linewidth]{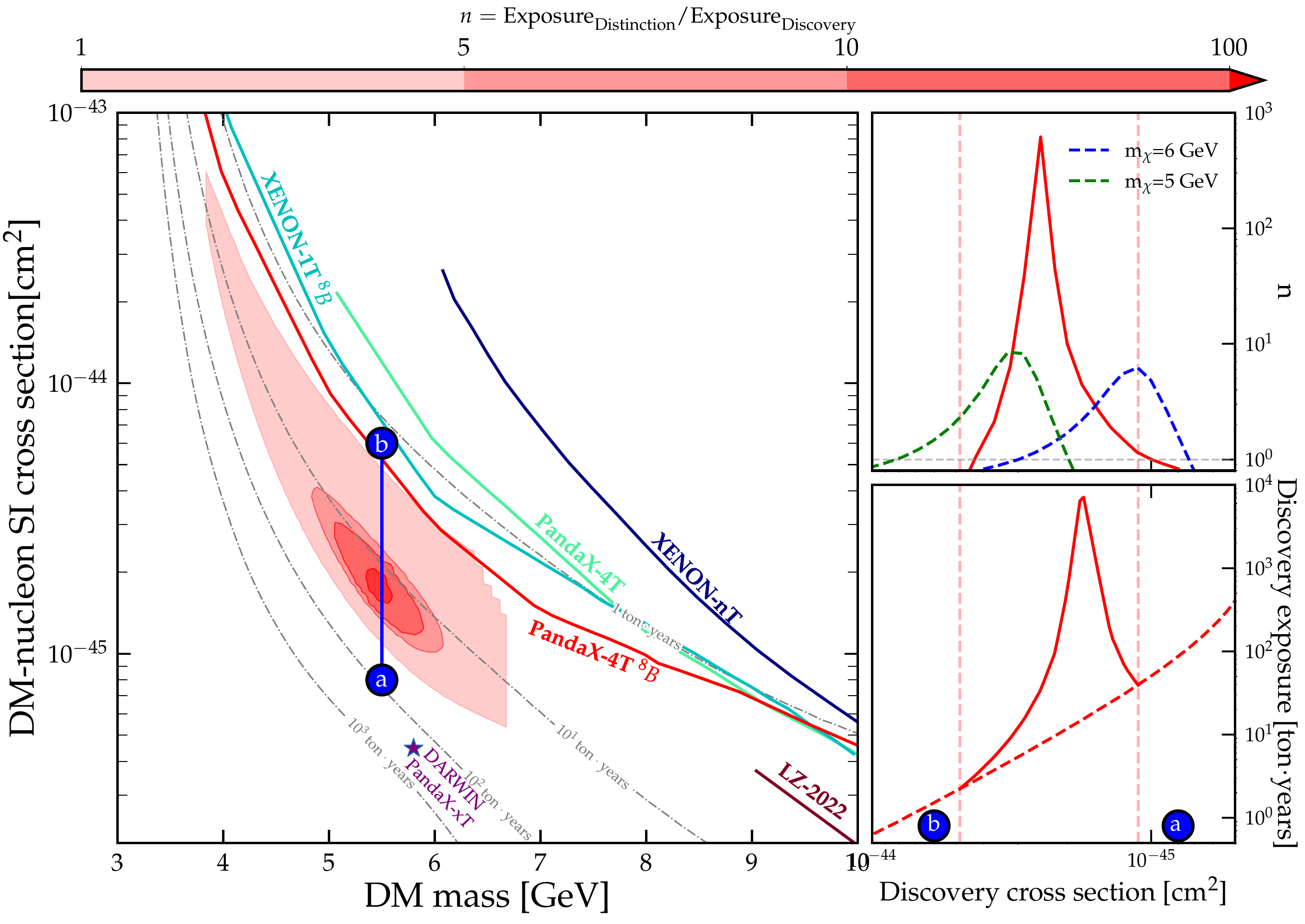}
	\caption{\label{fig:DMN-SMvsDMNP}The distinction plot for $\rm DM+\nu NP~v.s.~SM$ and DM-nucleon interaction, together with experimental exclusion limits~\cite{Liu:2022zgu,LZ:2022lsv,PandaX:2022aac,XENON:2020gfr,XENON:2023cxc} presented by colored solid lines. \zbl{Here, $\nu$NP refers to the \UUniversal model, with $\{m_{Z^\prime},~g_\mathrm{Universal}\}=\{30~\mathrm{MeV},~2\times 10^{-5}\}$.}}
\end{figure}

Now we intend to explain how to present the quantitative results in the simultaneous presence of DM and $\nu$NP represented by $\rm DM+\nu NP~v.s.~SM$, and derive interesting results from the mock data analyses.
Based on the discovery limit, we can establish a correlation between the DM-nucleon cross section and the requested discovery exposure, as depicted by the dashed-red line in the lower right panel of Fig.~\ref{fig:DMN-SMvsDMNP}, where $m_\chi = 5.5~\mathrm{GeV}$ is assumed. 
When considering the presence of $\nu$NP, specifically the \UUniversal model, the spectra of two hypotheses become quite similar for a certain range of cross sections. As the exposure increases, the distinction power must be stronger in statistics. Consequently, the experiment requires a higher exposure to observe DM and $\nu$NP signals, as depicted by the red line in the lower right panel of Fig.~\ref{fig:DMN-SMvsDMNP}. Their ratio $n$ is illustrated as the red line in the upper right panel of Fig.~\ref{fig:DMN-SMvsDMNP}.
For $m_\chi = 5.5~\mathrm{GeV}$, the cross section range for $n\geq 1$ is approximately $1.1\times10^{-45}~\mathrm{cm}^2 \lesssim \sigma_{\chi-p}^0 \lesssim 4.6\times10^{-45}~\mathrm{cm}^2$. The corresponding discovery exposures are $3.9~\mathrm{ton\times years}$ and $0.3~\mathrm{ton\times years}$, respectively. Note that we assume a constant NR detection efficiency of $0.1$ according to Ref.~\cite{PandaX:2022aac,XENON:2020gfr}. At $\sigma_{\chi-p}^0 = 1.8\times10^{-45}~\mathrm{cm}^2$ with $n= 617.6$, the maximum point implies that the experiment needs to increase the discovery exposure by a factor of about $600$ in order to achieve the distinction goal. \zbl{Numerically, the discovery and distinction exposures at this point are $921.3~\mathrm{ton\times years}$ and $1.5~\mathrm{ton\times years}$, respectively. }
In the upper right panel of Fig.~\ref{fig:DMN-SMvsDMNP}, we also depict the ratio $n$ as a function of the discovery cross section for different DM masses. These ratios exhibit different shapes because DM with varying masses generates different spectra as shown in Fig.~\ref{fig:ERSpectrum}. DM with a lower mass produces fewer events due to kinematic limitations so that it requires a larger cross section to make a discovery given by a fixed exposure. Consequently, for $m_\chi = 5~\mathrm{GeV}$, the distinction area shifts to the regions with larger cross sections.

As shown in the left panel of Fig.~\ref{fig:DMN-SMvsDMNP}, we illustrate the area where the presence of DM and $\nu$NP can not be distinguished from \zbl{Standard Model CE$\nu$NS} with the DM discovery exposure. In this region, an experiment might not observe any excess from CE$\nu$NS backgrounds, despite DM really exists. Due to the presence of $\nu$NP, \zbl{the excesses in DM signals within these regions compensate for of the depletion of neutrino backgrounds.} The most opaque region is around $m_\chi = 5.5~\mathrm{GeV}$ and $\sigma_{\chi-p}^0 = 1.8\times 10^{-45}~\mathrm{cm}^2$. 
Outside the shaded area, the experiment can readily exclude the absence of both DM and $\nu$NP given the DM discovery exposure.
It is noted that the distinction regions for \UUniversal can be completely examined by future experiments DARWIN~\cite{DARWIN:2016hyl} and PandaX-xT~\cite{PandaX:2024oxq} with a projected exposure of $200~\mathrm{ton\times years}$. \zbl{Here, we assume an energy threshold of 1~keV and a constant NR detection efficiency of 0.1. Future experiments will have an exposure that is two orders of magnitude larger than that of current experiments.}
What stands out in Fig.~\ref{fig:DMN-SMvsDMNP} is that the red region almost intersects with the DM excluded parameter space. This fact tells us that the existence of DM in the excluded region might still be allowed given the presence of \UUniversal. The potential and considerable effect of $\nu$NP on the current exclusion limits will be discussed in Part~\ref{sec:exclusion}.

\begin{figure}[htbp]
	\centering
	\includegraphics[width=0.7\linewidth]{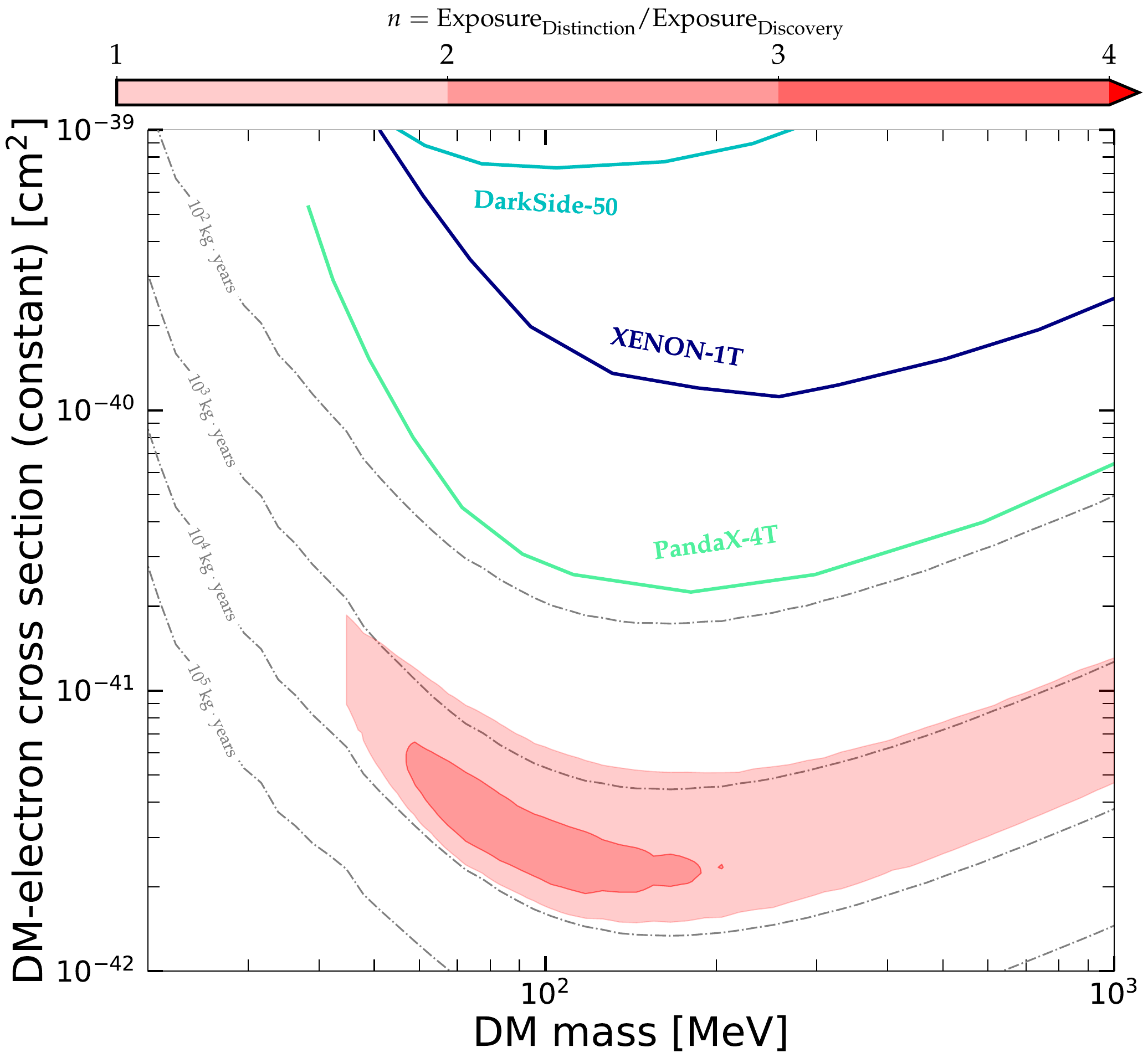}
	\caption{\label{fig:DMe-SMvsDMNP}The distinction plot for $\rm DM+\nu NP~v.s.~SM$ and DM-electron interactions.}
\end{figure}
We are also interested in exploring the context of DM-electron interactions, which has been less discussed in the literature before. As shown in Fig.~\ref{fig:DMe-SMvsDMNP}, we present the distinction area over the parameter space of DM-electron cross section v.s. DM mass. 
For $m_\chi = 100~\mathrm{MeV}$, \UUniversal affects the DM-electron cross section $\sigma_{\chi-e}^0$ ranging from $1.7\times10^{-42}~\mathrm{cm}^2 $ to  $5.3\times10^{-42}~\mathrm{cm}^2$. The corresponding discovery exposures are $881.0~\mathrm{kg\times years}$ and $110.3~\mathrm{kg\times years}$, respectively. Note that we assume a constant ER detection efficiency of $0.1$ according to Ref.~\cite{PandaX:2022xqx,XENON:2019gfn}. At the point $\sigma_{\chi-e}^0 = 2.4\times10^{-42}~\mathrm{cm}^2$, an exposure about three times larger than the discovery exposure is required to observe the DM and $\nu$NP signals.
The distinction area is still far away from the current limits in the PandaX-4T experiment. However, the presence of $\nu$NP might be carefully considered for the future experiment with a higher exposure and a better sensitivity.
Moreover, it is worth noting $\nu$NP has fewer impacts in the context of DM-electron interactions than that in the context of DM-nucleon interactions, which can be seen from the strength of the ratio $n$. This phenomenon can be understood by a glimpse of the greater difference between DM and neutrino spectra for DM-electron interactions.

\subsection{$\rm DM~v.s.~\nu NP$}
\label{sec:B}
\begin{figure}[htbp]
	\centering
	 \includegraphics[width=0.48\linewidth]{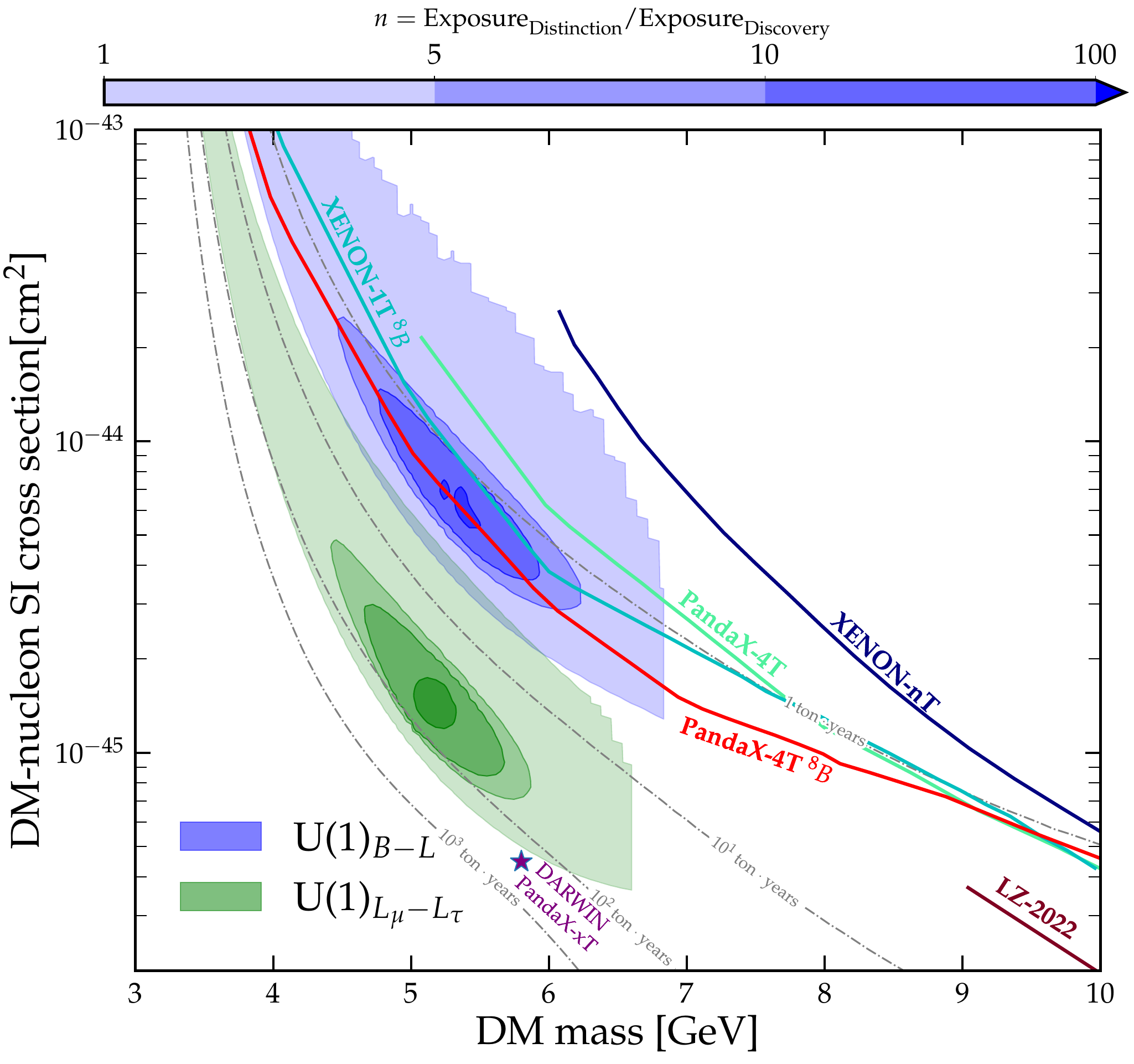}
  \includegraphics[width=0.48\linewidth]{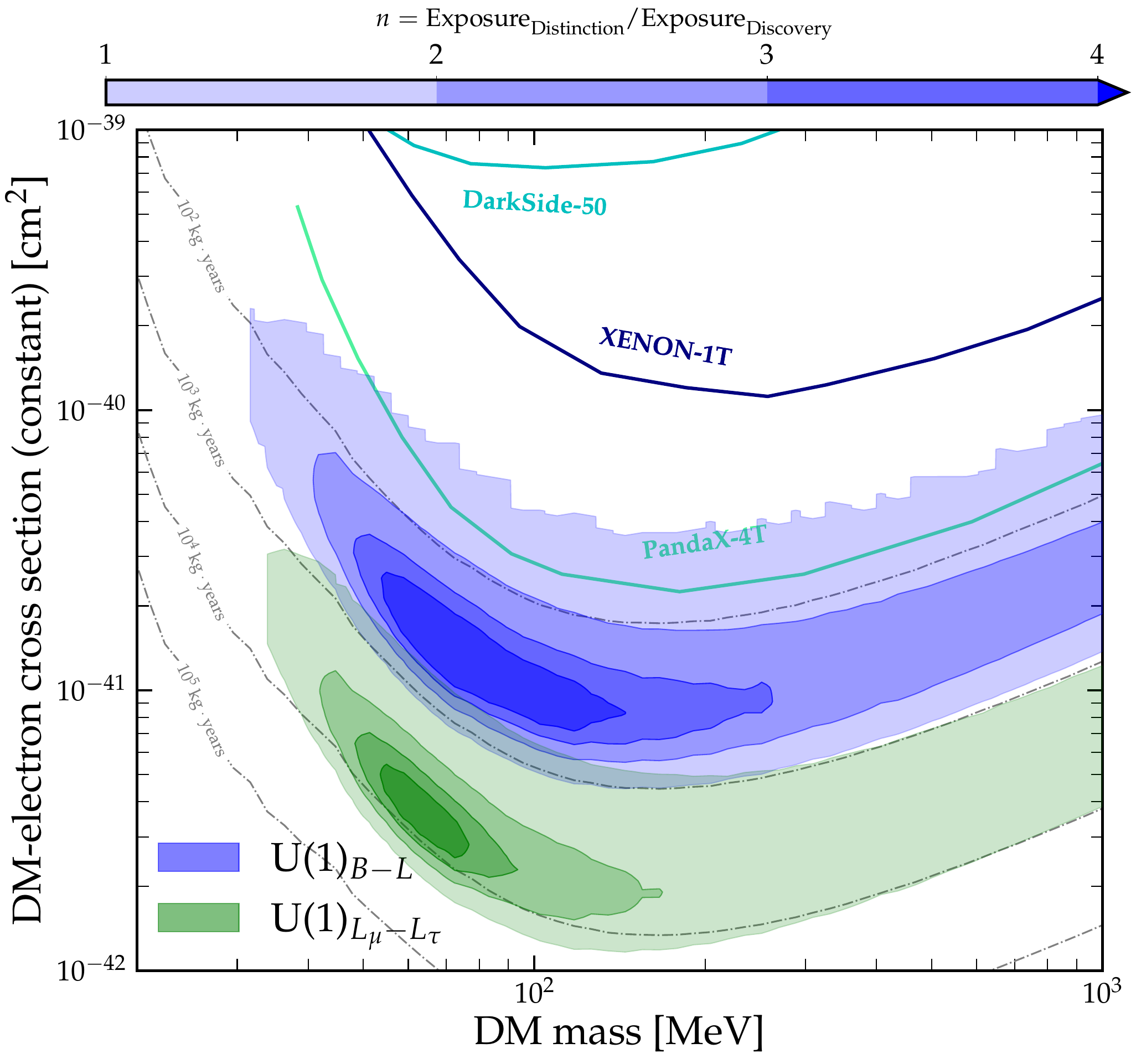}
	\caption{\label{fig:DMvsNP}The $\rm DM~v.s.~\nu NP$ distinction plots for DM-nucleon interactions (left) and DM-electron interactions (right). \zbl{The parameters for the \UBL and \ULTau model adopted here are $\{m_{Z^\prime},~g_\mathrm{B-L}\}=\{20~\mathrm{MeV},~4\times 10^{-5}\}$, $\{m_{Z^\prime},~g_\mathrm{L_\mu-L_\tau}\}=\{16~\mathrm{MeV},~8\times 10^{-4}\}$, respectively.}}
\end{figure}
This part seeks to analyse the quantitative results for the scenario $\rm DM~v.s.~\nu NP$, where a competition of DM and $\nu$NP exists.
It was already shown that either \UBL or \ULTau would offer an excess of neutrino events compared with SM predictions in the DM detector, where \UBL exhibits a more pronounced deviation from CE$\nu$NS when specific benchmark points are considered. We examine the statistical distinction of DM and $\nu$NP in contexts of DM-nucleon and DM-electron interactions in the Fig.~\ref{fig:DMvsNP}. The shaded regions in both models point to the fact that, given the original DM discovery exposure, an experiments can not determine whether the potential excess originates from DM or $\nu$NP. 
The distinction regions for two models are obviously different, because additional neutrino spectra from $\nu$NP mimic the spectra generated by different DM. 
As shown in the left panel of Fig.~\ref{fig:DMvsNP}, in the context of DM-nucleon interactions, the most opaque region for \UBL is around $m_\chi = 5.4~\mathrm{GeV}$ and $\sigma_{\chi-p}^0 = 6.7 \times 10^{-45}~\mathrm{cm}^2$, where $n= 189.7$. While for \ULTau  the most opaque region is around $m_\chi = 5.2~\mathrm{GeV}$ and $\sigma_{\chi-p}^0 = 1.5 \times 10^{-45}~\mathrm{cm}^2$, where $n= 280.3$.
Interestingly, the blue region encroaches upon the DM excluded parameter space. From the statistical point of view on exclusion limits, the excluded parameter space corresponds to an allowed fluctuation on observed spectra. Therefore, if the allowed fluctuation is assigned to the presence of \UBL, \zbl{there will be more stringent constraints on DM.} This effect will be further discussed in Part~\ref{sec:exclusion}.
\zbl{While \ULTau exhibits little influence in current experiments,} it will have significant impacts on DM searches in future experiments, such as DARWIN and PandaX-xT with a projected exposure of $200~\mathrm{ton\times years}$.

In the right panel of Fig.~\ref{fig:DMvsNP} , it can be seen that in the context of DM-electron interactions, the most opaque region for \UBL is around $m_\chi = 72~\mathrm{MeV}$ and $\sigma_{\chi-e}^0 = 1.4 \times 10^{-41}~\mathrm{cm}^2$, where $n= 6.3$. As for \ULTau, the most opaque region is around $m_\chi = 63~\mathrm{MeV}$ and $\sigma_{\chi-e}^0 = 3.6 \times 10^{-42}~\mathrm{cm}^2$, where $n= 5.2$.
It is noteworthy that for some specific models like \UBL, the excluded parameter space touches the shaped region, though not manifestly. This suggests that in the near future, greater attention should be paid to distinguishing between DM and $\nu$NP signals.

\subsection{Impacts on DM Exclusion Limits}
\label{sec:exclusion}

As discussed above, the distinction regions have invaded the current excluded DM parameter space in the context of DM-nucleon interactions. In the following, we take PandaX-4T as an example to demonstrate how the presence of U(1) extensions alters the exclusion limits.
Though the XENON and PandaX collaborations have not observed CE$\nu$NS events originating from solar neutrinos~\cite{XENON:2020gfr, PandaX:2022aac}, they have presented provided 90\% C.L. upper limits on the $^8$B solar neutrino flux: \zbl{$1.4\times 10^{7}~\rm{cm^{-2}s^{-1}}$ for XENON1T with an exposure of $0.6~\mathrm{ton\times years}$, and $9\times 10^{6}~\rm{cm^{-2}s^{-1}}$ for PandaX-4T with an exposure of $0.48~\mathrm{ton\times years}$, respectively.} As a reference, the measured $^8$B neutrino flux by the dedicated solar neutrino experiment is approximately $5.46\times 10^{6}~\rm{cm^{-2}s^{-1}}$. Furthermore, they have also updated more restrictive exclusion limits on DM with new data analysis techniques. 
\begin{figure}[htbp]
	\centering
	\includegraphics[width=0.6\linewidth]{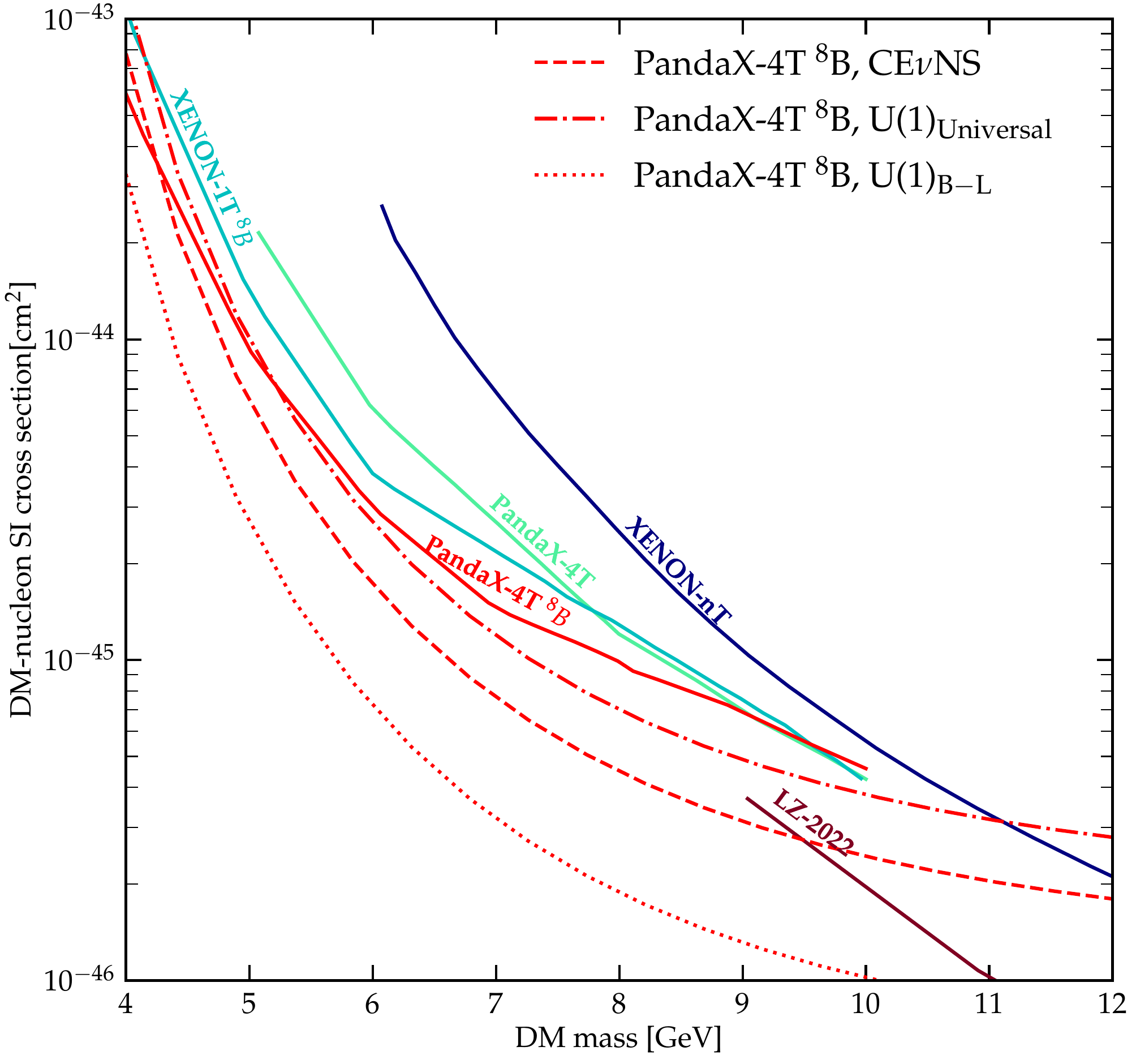}
	\caption{\label{fig:exclusions}The 90\% C.L. exclusion limits for DM-nucleon interaction. Solid lines are taken from DM direct detection experiments.}
\end{figure}

For the sake of simplicity, we derive our exclusion limits on DM by simply transforming the upper limits on the $^8$B solar neutrino flux. For PandaX-4T, the region of interest (ROI) is set to $\rm 1~keV\ 3~keV$ according to Ref.~\cite{PandaX:2022aac}. Here we set the energy threshold at $1~\mathrm{keV}$ due to the significant suppression on efficiencies for nuclear recoil energies $\lesssim~1~\mathrm{keV}$~\footnote{\zbl{Both Xenon-1T~\cite{XENON:2020gfr} and PandaX-4T~\cite{PandaX:2022aac} have analysis thresholds below 1~keV. For the purpose of simplicity, in this article, we adopt 1~keV as the energy threshold and derive the NR detection efficiency from Ref.~\cite{PandaX:2022aac}.}}. The total CE$\nu$NS events in this region is approximately $n_\mathrm{CE\nu NS}=1.66$ with an exposure of $0.48~\mathrm{ton\times years}$, where the NR detection efficiency was also taken into account. Denoting the upper limit event by $n_\mathrm{limit}=2.74$, the exclusion limit for the DM-nucleon cross section is determined by:
\begin{equation}\label{eqn:exclusion}
n_\mathrm{DM}(\sigma_{\chi-p}^0) = n_\mathrm{limit} - n_{\nu}\,,
\end{equation}
where $n_\mathrm{DM} \propto \sigma_{\chi-p}^0$ and $n_{\nu}$ are the DM and neutrino events in ROI, respectively. Here we have $n_{\nu}=n_\mathrm{CE\nu NS}$ for the \zbl{SM} predicted interaction only, as the exclusion limits shown in Fig.~\ref{fig:exclusions}. Please note that $n_{\nu}$ is modified when $\nu$NP is considered. 
As shown in Fig.~\ref{fig:exclusions}, the red dashed line represents our exclusion limits recasted from constraints on the $^8$B solar neutrino flux in PandaX-4T. We see the clear difference between our limits and those of PandaX-4T. Here we emphasize that our limits are oversimplified only to motivate the relative impact of $\nu$NP on DM exclusion limits while the dedicated analyses are still needed in experiments.

Referring to~Eqn.~(\ref{eqn:exclusion}), $n_\mathrm{DM}$ becomes larger if $n_{\nu}$ is suppressed by $\nu$NP. Consequently, a reduction of neutrino events caused by $\nu$NP leads to an increase of the excluded DM-nucleon cross section, and vice versa.
As the red dashed-dotted line shown in Fig.~\ref{fig:exclusions}, the \zbl{$\nu$NP} like \UUniversal looses the exclusion limits on DM. On the contrary, the exclusion limits are suppressed by an enhancement of neutrino events induced by $\nu$NP such as \UBL. Note that the coupling constant has been adjusted to $g_\mathrm{B-L}=2\times 10^{-5}$, because results in the original choice of $n_{\nu}$ larger than $n_\mathrm{limit}$. These observations confirm our discussion in Part~\ref{sec:A} and Part~\ref{sec:B}, where we emphasize that the influence of $\nu$NP on DM excluded limits is considerable.
Although our results are derived from simulated data without full-fledged analyses, they effectively imply the necessity to consider $\nu$NP in DM exclusion limits.

\subsection{\zbl{Results with different energy thresholds}}
\label{sec:D}
\begin{figure}[htbp]
	\centering
	\includegraphics[width=0.9\linewidth]{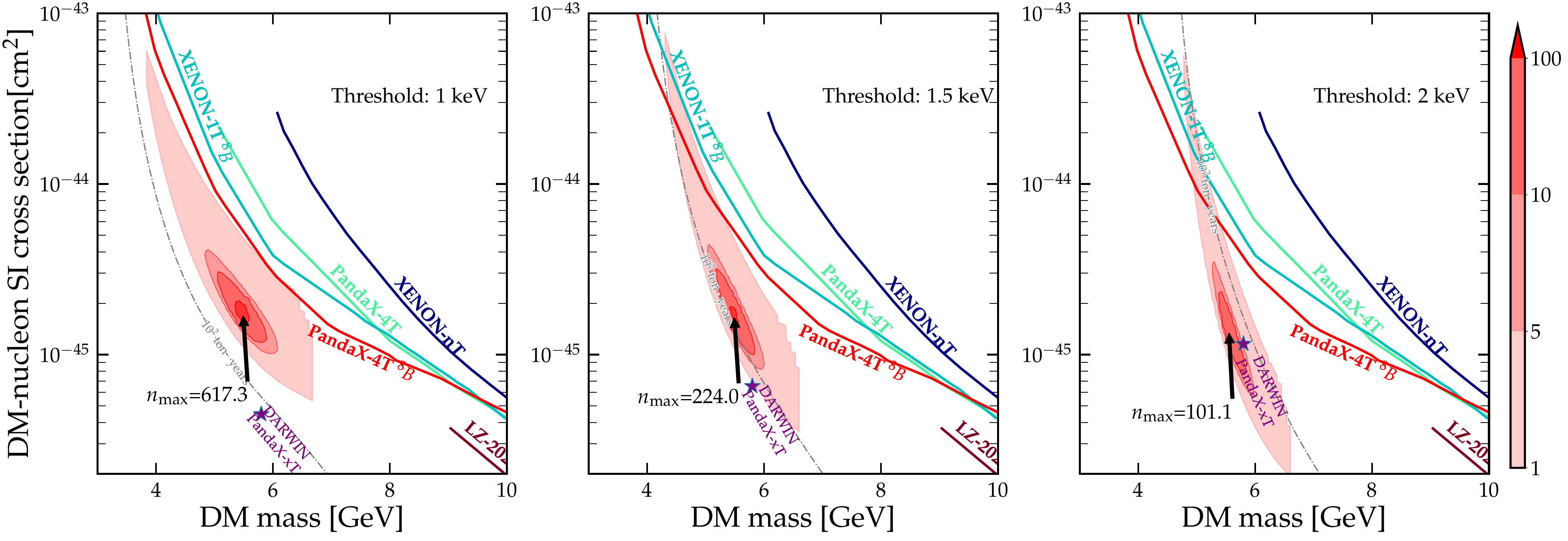}
	\caption{\label{fig:DMN_Threshold_1} \zbl{The distinction plot for $\rm DM+\nu NP~v.s.~SM$ and DM-nucleon interaction given different detector thresholds.} }
\end{figure}

\zbl{In this part, we delve into the implications of varying the energy threshold. For brevity, we initially focus on the $\rm DM+\nu NP~v.s.~SM$ case, while we leave the results for $\rm DM~v.s.~\nu NP$ case in the Appendix~\ref{appendix4}. 
The spectra of DM and neutrino undergo modifications depending on varying thresholds. In turn, we see a varying discrepancy between spectra under two statistical hypotheses. Therefore, as shown in Fig.~\ref{fig:DMN_Threshold_1}, the distinction regions for the DM-nucleon interaction become narrower and shallower with an increasing energy threshold. This trend suggests that a higher threshold leads to a more significant discrepancy between spectra obtained under two statistical hypotheses. However, a higher threshold also correlates with lower event rates, necessitating a greater exposure to detect DM signals, regardless of the presence of $\nu$NP. Fig.~\ref{fig:DMN_Threshold_1} illustrates that a higher threshold restricts the detection of weaker DM-nucleon interactions, given a fixed exposure.
For the $\rm DM~v.s.~\nu NP$ case, please see Appendix~\ref{appendix4} where the results are similar. 
Consequently, with the advent of low energy threshold detection technologies and significant advances in statistical data analysis, experimental physicists should still exercise caution in accounting for the influence of $\nu$NP.}

\begin{figure}[htbp]
	\centering
	\includegraphics[width=0.9\linewidth]{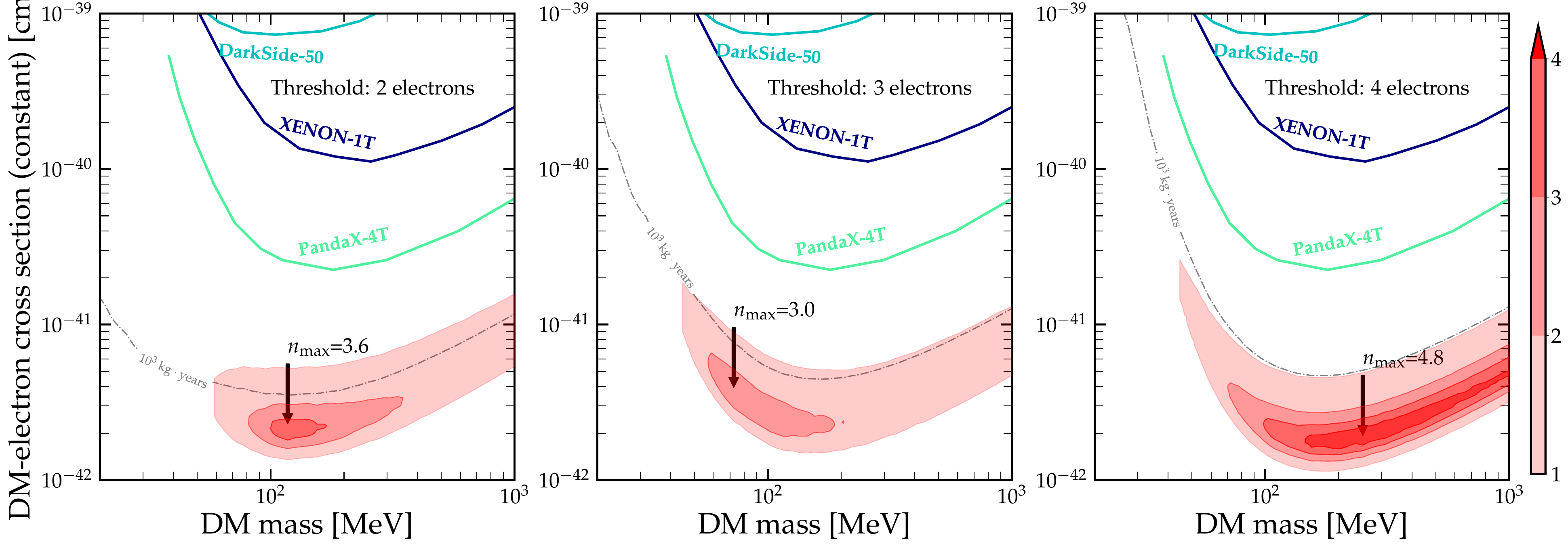}
	\caption{\label{fig:DMe_Threshold_1} \zbl{The distinction plot for $\rm DM+\nu NP~v.s.~SM$ and DM-electron interaction given different detector thresholds.}}
\end{figure}

\zbl{As shown in Fig.~\ref{fig:DMe_Threshold_1}, when compared to a threshold set at four emitted electrons, a threshold of three emitted electrons mitigates the impact of $\nu$NP for the DM-electron interaction scenario for certain parameter regions. By pinning down a threshold of three emitted electrons, an experiment not only gains more statistics, but also amplifies the discrepancy between spectra obtained under two hypotheses. This is opposite to the DM-nucleon interaction scenario, exhibiting a notable shift of the saturated regions. However, lowering the threshold from three to two emitted electrons leads to a more degenerate spectra.
Likewise, the findings and conclusions are quite similar to the $\rm DM~v.s.~\nu NP$ case and the \UBL model, please see Appendix~\ref{appendix4}. The disparity in the results for \ULTau compared to other models indicates that aforementioned conclusions are model-dependent.
Therefore, it deserves efforts to improve the detector threshold in DM experiments to quantify the impact of $\nu$NP for the DM-electron interaction scenarios.}

\subsection{Model interpretations}
\label{sec:E}
\begin{figure}[htbp]
	\centering
  \includegraphics[width=0.48\linewidth]{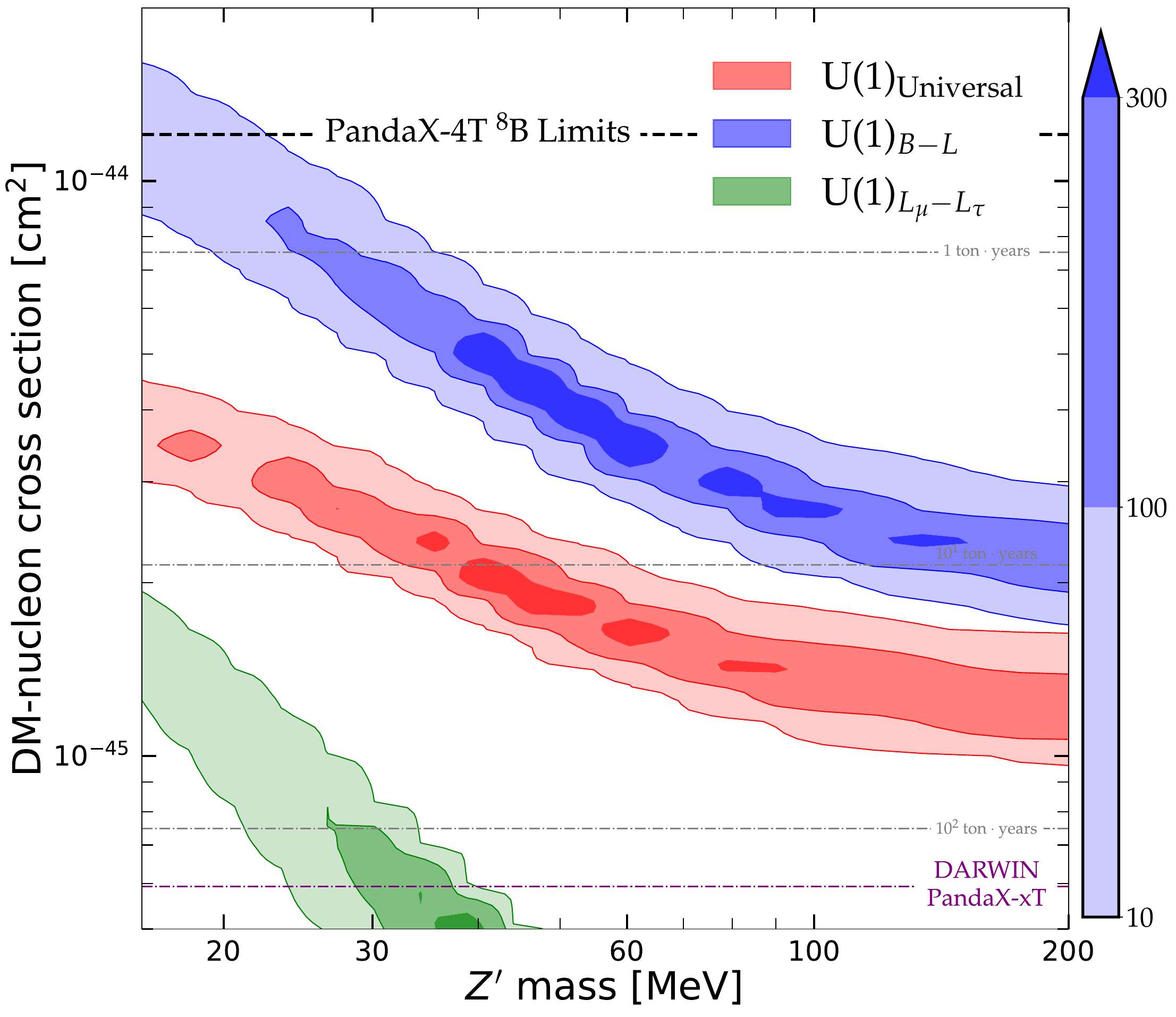}
	\includegraphics[width=0.48\linewidth]{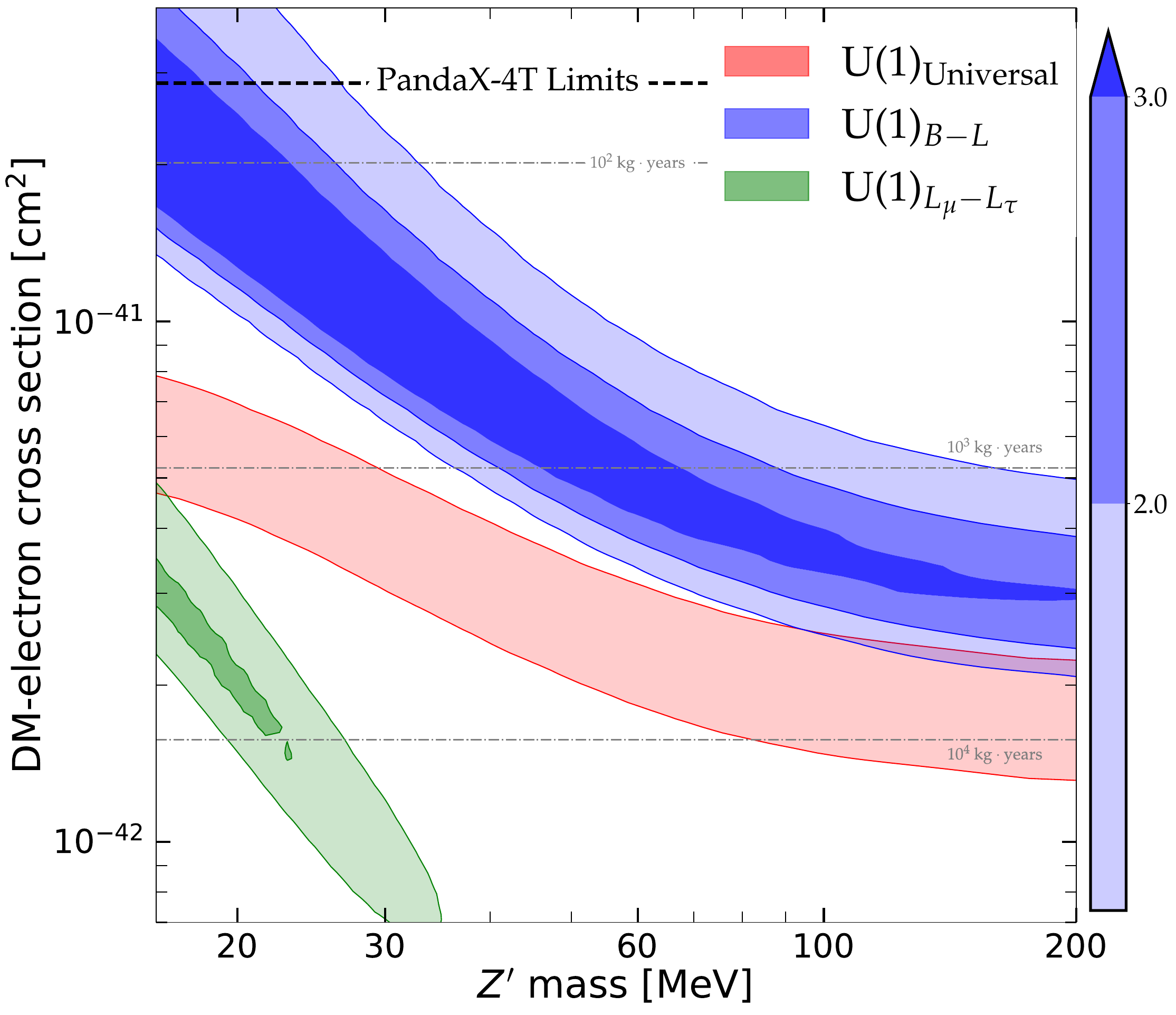}
	\caption{\label{fig:DMU1}The distinction plots on the planes of the DM-nucleon cross section v.s. $m_{Z^\prime}$  and the DM-electron cross section v.s. $m_{Z^\prime}$ for different U(1) models. Here, $m_\chi=5.5~\mathrm{GeV}$ in the left panel, while $m_\chi=100~\mathrm{MeV}$ in the right panel. }
\end{figure}

This part attempts to comprehensively present the effects of U(1) models on DM searches by exploring different model parameters. \zblt{As illustrated in Fig.~\ref{fig:U1-constraints}, these parameters are selected from the maximally allowed region given in Ref.~\cite{AtzoriCorona:2022moj}, ensuring that the U(1) coupling constant is fixed if $m_{Z^\prime}$ is determined.} Note that we only consider the mass range $\rm 16~\mathrm{MeV}\lesssim m_{Z^\prime}\lesssim 200~\mathrm{MeV}$, since models within this range do not get much constrained yet. Due to the low energy threshold of DM detectors, light mediators can significantly modify neutrino spectra.
Specifically, we use $m_\chi=5.5~\mathrm{GeV}$ for DM-nucleon interactions and $m_\chi=100~\mathrm{MeV}$ for DM-electron interactions as benchmark points. This allows us to project the ratio of the distinction and discovery exposure $n$ onto the DM-nucleon and DM-electron cross sections versus $m_{Z^\prime}$ plane, as shown in Fig.~\ref{fig:DMU1}.

\zbl{The constraints on \UUniversal and \UBL are derived from COHERENT experiments~\cite{COHERENT:2017ipa,COHERENT:2020iec},} which have higher thresholds than DM direct detection experiments. The presence of light mediators produces more events in the low recoil energy, such as those observed in the DM detectors. This indicates the complementarity of CE$\nu$NS measurements between DM experiments and artificial neutrino source experiments. Under the constraints from COHERENT, the lighter mediators lead to more significant modifications in neutrino spectra within DM detectors. Therefore, an increase of $m_{Z^\prime}$ results in a reduction of the cross section for the distinction limit as shown in Fig.~\ref{fig:DMU1}. Likewise, the discussion for \ULTau is similar, \zbl{except that constraints on \ULTau mainly come from analyses of neutrino trident production processes~\cite{CCFR:1991lpl,Altmannshofer:2014pba}.}
One can see from Fig.~\ref{fig:DMU1} that U(1) models have more significant impacts in the context of DM-nucleon interactions than those in DM-electron interactions, due to the degenerate spectra in the two distinction scenarios. In terms of DM-nucleon interactions, most U(1) models require an exposure with two orders of magnitude larger than the DM discovery exposure. 
Moreover, current limits have reached the distinction area for \UBL model ($n\geq 10$ and $n\geq 2$ for DM-nucleon and DM-electron interactions, respectively), indicating that effects of \UBL with $m_{Z^\prime}\lesssim 40~\mathrm{MeV}$ on DM searches are sizable. One can see from the left panel of Fig.\ref{fig:DMU1} that in future experiments DARWIN and PandaX-xT, the influence of \UUniversal and \ULTau would be also considerable. Therefore, U(1) extensions with a lighter mediator deserve more careful consideration in the current and future DM experiments.

\section{Conclusion and outlook}
\label{sec:conclusion}
In the present study, the reconstructed spectra from a mixture of events associated with DM, CE$\nu$NS, and \zbl{$\nu$NP} were investigated in a Xenon-based DM experiment. To better understand the impact of $\nu$NP on DM searches, the concept of the distinction limit was introduced for quantitative analyses. The calculation method for determining the distinction limit is efficient and available in an open-source repository~\cite{ourcode}, which might facilitate analyses in DM direct detection experiments and support other phenomenological research.

The study demonstrated the impact of $\nu$NP on DM direct searches with mock data for recoil spectra, incorporating both DM-nucleon and DM-electron interactions.
Considering U(1) extensions with fixed benchmark points, it was found that \UUniversal, which depletes neutrino event rates, could lead to DM signals being masked by neutrino backgrounds, potentially relaxing constraints on the traditional DM exclusion region. 
Conversely, U(1) extensions \UBL and \ULTau, which enhance neutrino event rates, could complicate the determination of whether excess signals are due to DM or $\nu$NP, potentially tightening the DM exclusion limits.
Additionally, the excluded DM regions have intersected with the distinction regions between DM and $\nu$NP, particularly for DM-nucleon interactions. It is recommended that DM direct detection collaborations adopt this analytical approach and carefully consider the effects of $\nu$NP, such as those from U(1) extensions, when determining DM exclusion limits. 
Further, improving the energy threshold of detectors may increase the statistic and the discrepancy of spectra simultaneously, potentially mitigating the impact of $\nu$NP, especially in scenarios involving DM-electron interactions and certain $\nu$NP models. Thus, optimizing detector thresholds is crucial in DM experiments. 
The analysis also indicated that U(1) models with lighter mediators have a more pronounced effect on DM searches in both current and future experiments. Notably, U(1) models have a more significant impact on DM-nucleon spin-independent interactions compared to DM-electron interactions mediated by heavy vector bosons.

To mitigate the influence of $\nu$NP on DM searches, several strategies to enhance the distinction in spectra warrant further exploration: i) incorporating timing information, such as the daily or annual modulation; ii) utilizing the angular information for the directional detection. 
Additionally, $\nu$NP constraints may be improved through other approaches, including neutrino oscillation studies, collider experiments, and beam dump experiments. The potential challenges posed by $\nu$NP in current and future Xenon-based DM experiments may be addressed in conjunction with these methods. 
The present study was limited to DM-nucleon and DM-electron interactions, as well as three U(1) extensions. Further research is recommended using similar statistical approaches to investigate DM-coupled effective operators, model-independent analyses of $\nu$NP, or other non-standard neutrino physics such as the neutrino magnetic moment and neutrino generalized interactions.
\zblt{Finally, we assume the weakly-interacting massive particles (WIMPs) as the DM candidate in this work. However, scenarios involving other light DM candidates, such as dark photon, axion, axion-like particle and fermionic DM, also deserve further investigation, where light DM candidates generate electron recoil events in DM detectors through dark absorption processes.}

\paragraph{Note added.} Recently, PandaX and XENON collaborations have reported their first observation of solar $^8$B neutrinos in Xenon-based DM experiments~\cite{PandaX:2024muv, XENON:2024ijk}. Our work highlighted the possibility of ``one stone with two birds": searching for DM
and $\nu$NP at the same time. 

\acknowledgments

We appreciate Qing Lin, Gang Li, Ning-Qiang Song and Jiang-Hao Yu for fruitful discussions. This project was supported in part by National Natural Science Foundation of China under Grant Nos. 12347105 and 12075326. This work was supported in part by Fundamental Research Funds for the Central Universities (23xkjc017) in Sun Yat-sen University. JT is grateful to Southern Center for Nuclear-Science Theory (SCNT) at Institute of Modern Physics in Chinese Academy of Sciences for hospitality.

\appendix
\section{The asymptotic formula of the test statistic}
\label{appendix1}
Following the same strategy in Ref.~\cite{Tang:2023xub}, the log-likelihood ratio $-2\ln \lambda(\theta^0_1,\theta^0_2)$ (and  $-2\ln \lambda(\theta^1_1, \theta^1_2))$ with a factor of $-2$ can be written asymptotically as:
\begin{equation}
-2\ln \lambda(\theta^0_1,\theta^0_2) \approx (\dot{l}(\boldsymbol{\theta}^\prime)+\mathscr{F} \boldsymbol{\delta}^0)^{T}[\mathscr{F}^{-1}-\mathbf{H}](\dot{l}(\boldsymbol{\theta}^\prime)+\mathscr{F} \boldsymbol{\delta}^0)\,,
\end{equation}
where  $\dot{l}(\boldsymbol{\theta}^\prime)$ is the first derivative of the logarithmic likelihood function $l(\boldsymbol{\theta}^\prime)$ at $\boldsymbol{\theta}^\prime$. The true value $\boldsymbol{\theta}^\prime$ can be taken as  $\boldsymbol{\theta}^0$ or $\boldsymbol{\theta}^1$ according to whether H$_0$ or H$_1$ is real, and we have $\boldsymbol{\delta}^0=\boldsymbol{\theta}^\prime-\boldsymbol{\theta}^0$. Note that $\dot{l}(\boldsymbol{\theta}^\prime) \sim \mathscr{N}(\boldsymbol{\mu}, \mathbf{V})$ follows a multivariate normal distribution with a mean vector $\boldsymbol{\mu}$ and a covariance matrix $\mathbf{V}$. $\mathscr{F} \equiv -E(\ddot{l}(\boldsymbol{\theta}^\prime))$ is the expectation of the second derivative of $l(\boldsymbol{\theta}^\prime)$ at $\boldsymbol{\theta}^\prime$. 
Explicitly,
$$
\mathscr{F}\equiv\left[\begin{array}{cc}
\mathbf{G}_1 (r \times r) & \mathbf{G}_2 (r \times(k-r))\\
\mathbf{G}_2^T((k-r) \times r) & \mathbf{G}_3((k-r) \times(k-r))
\end{array}\right], \quad
\mathbf{H}\equiv\left[\begin{array}{cc}
0 & \mathbf{0} \\
\mathbf{0} & \mathbf{G}_3^{-1}
\end{array}\right] \,.
$$
where $\mathbf{G}_1,~\mathbf{G}_2,~\mathbf{G}_3$ are the block matrices inside $\mathscr{F}$, and $r,~k$ represent their dimensions. Note that $k$ and $r$ ($1\leq r \leq k$ ) are the number of all parameters and parameters of interest, respectively. In this paper, $r=2$ since we only care about the existence of DM and $\nu$NP, and the number of nuisance parameter $k-r=2$ since we only consider the uncertainties of the $^8$B and hep fluxes.

Therefore, the asymptotic formula for the test statistic is:
\begin{equation}
\begin{aligned}
q \approx & 2[(\boldsymbol{\delta}^0)^T -(\boldsymbol{\delta}^1)^T][1-\mathscr{F}\mathbf{H}] \mathbf{V}^{\frac{1}{2}}Z 
+ (\mathscr{F}\boldsymbol{\delta}^0)^T[\mathscr{F}^{-1}-\mathbf{H}](\mathscr{F}\boldsymbol{\delta}^0) - (\mathscr{F}\boldsymbol{\delta}^1)^T[\mathscr{F}^{-1}-\mathbf{H}](\mathscr{F}\boldsymbol{\delta}^1) \\
= & \mathbf{b} Z+c\,,
\end{aligned}
\end{equation}
where a vector $Z = \mathbf{V}^{-\frac{1}{2}} \dot{l}(\boldsymbol{\theta}^\prime)\sim \mathscr{N}(\mathbf{V}^{-\frac{1}{2}} \boldsymbol{\mu}, I)$ is introduced for convenience.
Therefore, the test statistic is the sum of several independent normal variance and constants, so it asymptotically follows a normal distribution: 
\begin{equation}
q \sim \mathscr{N}(\sum_{i}b_{i} \mu_i+c, \sqrt{\sum_{i}b_{i}^2})\,.
\end{equation}

The general expressions for $\boldsymbol{\mu}$, $\mathbf{V}$ and $\mathscr{F}$ can be found in Ref.~\cite{Tang:2023xub}.
In our case, since
\begin{equation*}
\begin{aligned}
&\frac{\partial v_i}{\partial \theta_j} = \left\{\begin{array}{cc}
    s_i\,, & j=1\,, \\
    \sum_j x_i^j \theta_j ~(j\geq 3), & j=2\,, \\
    \theta_2 x_i^j+b_i^j, & 3\leq j \leq M\,,
\end{array}\right.
\\
&\frac{\partial^2 v_i}{\partial \theta_j^2} =  0\,,
\end{aligned}
\end{equation*}
we have:
\begin{equation}\label{eqn:EandV}
\begin{aligned}
&\mu_\alpha = 0,\quad 1\leq \alpha \leq M\,,
\\
&\mathscr{F}_{\alpha \beta} = \left\{\begin{array}{cc}
\sum_{i} \frac{1}{v_i} \frac{\partial v_i}{\partial \theta_\alpha} \frac{\partial v_i}{\partial \theta_\beta} \,, & \alpha \in \{1,2\}~\mathrm{or}~\beta \in \{1,2\}\,, \\
\sum_{i} \frac{1}{v_i}\frac{\partial v_i}{\partial \theta_\alpha} \frac{\partial v_i}{\partial \theta_\beta} 
+\delta^\alpha_\beta \frac{1}{\sigma_{\alpha}^2}, & 3\leq i,j \leq M\,,
\end{array}\right.
\\
&V_{\alpha \beta} = \sum_{i j k} \left(\frac{1}{v_i} \delta^i_j
+\frac{\partial v_i}{\partial \theta_k} \frac{\partial v_j}{\partial \theta_k} \frac{\sigma_k^2}{v_i v_j}\right) 
\frac{\partial v_i}{\partial \theta_\alpha} \frac{\partial v_j}{\partial \theta_\beta},~1\leq \alpha,\beta \leq M\,,
\end{aligned}
\end{equation}
where $\delta^i_j$ is the Kronecker delta symbol.
Distinctive hypotheses will lead to different sets of parameters in a normal distribution. When $\mathrm{H}_0$ is true, the statistical measure $q$ will be negative. Otherwise, $q>0$ when $\mathrm{H}_1$ is true.

\section{Monte Carlo realisations}
\label{appendix2}
\begin{figure}[htbp]
	\centering
	\includegraphics[width=0.7\linewidth]{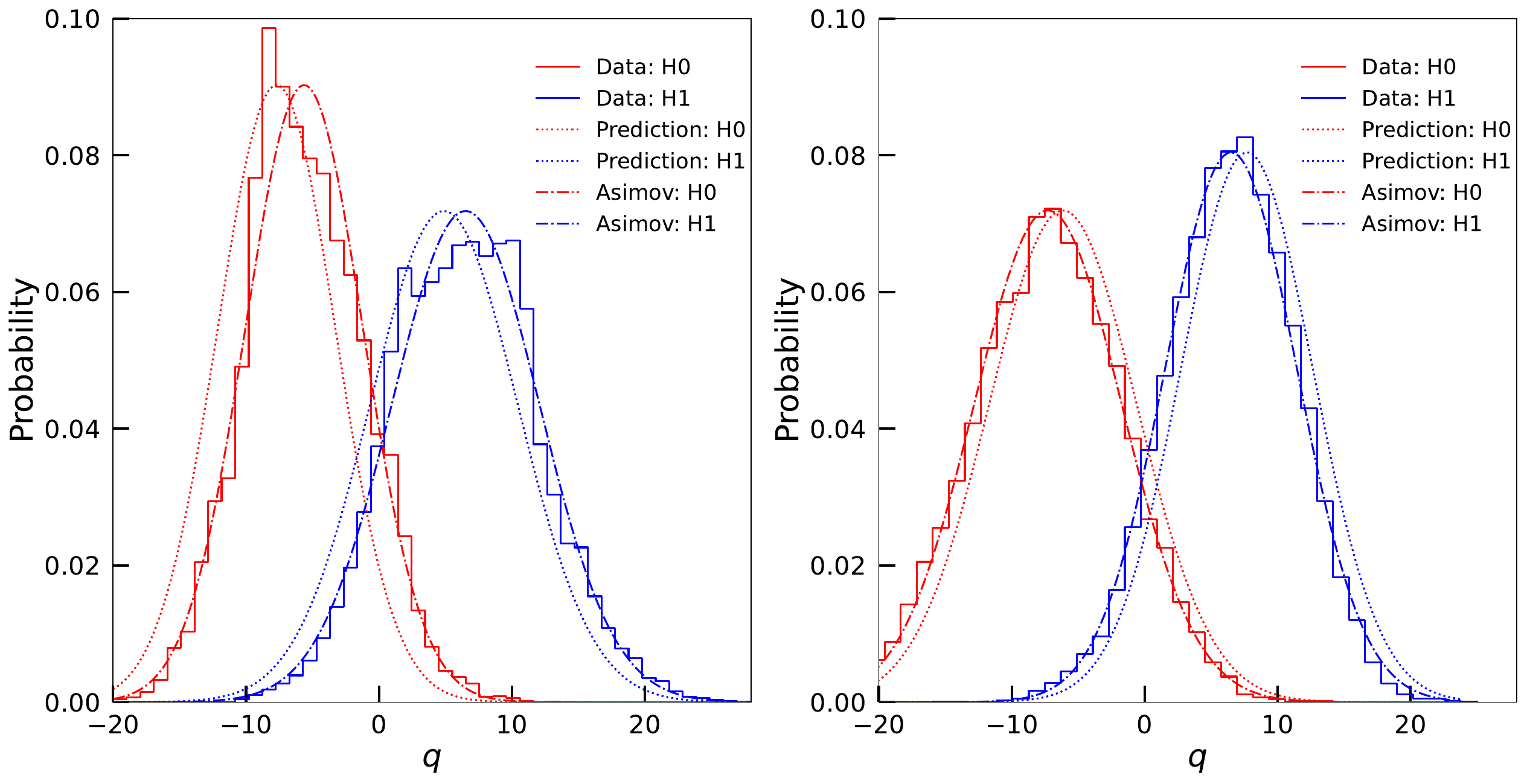}
	\caption{\label{fig:MC}Comparison between the test statistic's distribution from the theoretical prediction and the 10000 MC pseudo-experiments for a lower exposure (left) and a higher exposure (right). The benchmark points are chosen from the points at which the blue line intersects the colored region as shown in Fig.~\ref{fig:DMN-SMvsDMNP}, meaning $n=1$.}
\end{figure}
To demonstrate the effectiveness of our asymptotic method, we perform MC pseudo-experiments for comparison, as shown in Fig.~\ref{fig:MC}. Taking the case H$_0$ is real as an example, represented by the red lines, the means of the distributions derived from our asymptotic method have small deviations from those of the MC data. In contrast, the means obtained by the Asimov dataset align well with the data. However, we have to emphasize here that we exchange the standard derivations of two cases derived from our asymptotic method so that our predictions relatively accord with the data. This fact can be demonstrated for numerous cases, and further details can be found in our public codes~\cite{ourcode}. By adopting the symmetric definition of the distinction limit, we can guarantee the effectiveness of our method. 

\section{\zbl{The neutrino propagation}}
\label{appendix3}
\zbl{In this appendix, we briefly review the calculation of the neutrino propagation from Sun to Earth, following closely~Refs.~\cite{Gonzalez-Garcia:2013usa,AristizabalSierra:2017joc}. To obtain the fluxes of neutrinos with different flavors at Earth, we need to solve the evolution equation:
$$i\frac{d}{dr}|\boldsymbol{\nu}\rangle  =\left[    \frac{1}{2E_\nu}\boldsymbol{U}\;    \boldsymbol{H_\text{vac}}\;\boldsymbol{U}^\dagger    +    \boldsymbol{H_\text{mat}}  \right]|\boldsymbol{\nu}\rangle\,,$$
where $|\boldsymbol{\nu}\rangle^T= |\nu_e, \nu_\mu, \nu_\tau \rangle^T$ is the neutrino ﬂavor eigenstate basis, $r$ is the propagation path, $\boldsymbol{U}=\boldsymbol{U}(\theta_{23})\boldsymbol{U}(\theta_{13})\boldsymbol{U}(\theta_{12})$ is the leptonic mixing matrix, $\boldsymbol{H_\text{vac}}=\text{diag}(0,\Delta m_{21}^2,\Delta m_{31}^2)$ is the neutrino Hamitonian in vacuum, $\boldsymbol{H_\text{mat}}=\sqrt{2}G_Fn_e(r)\mathrm{diag}(1,0,0)$ with $n_e(r)$ the solar electron number density for SM. In the presence of of the neutrino non-standard interaction (NSI) or light U(1) models, the matter potential is changed so that the evolution equation becomes:
$$i\frac{d}{dr}|\boldsymbol{\nu}\rangle=\left[\frac{1}{2E_{\nu}}\boldsymbol{U}\boldsymbol{H}_{\mathrm{vac}}\boldsymbol{U}^{\dagger}+\sqrt{2}G_{F}n_{e}(r)\sum_{f=e,u,d}\boldsymbol{\varepsilon}^{\boldsymbol{f}}\right]|\boldsymbol{\nu}\rangle\,,$$
where $\boldsymbol{\varepsilon}^f=\begin{pmatrix}1+\varepsilon_{ee}^f&\varepsilon_{e\mu}^f&\varepsilon_{e\tau}^f\\\varepsilon_{e\mu}^f&\epsilon_{\mu\mu}^f&\varepsilon_{\mu\tau}^f\\\varepsilon_{e\tau}^f&\varepsilon_{\mu\tau}^f&\varepsilon_{\tau\tau}^f\end{pmatrix}$ with  $\varepsilon_{ij}^f(r)=Y_f(r)\epsilon_{ij}^f$ ($f=e,u,d$) and the relative abundance $Y_f(r)=n_f(r)/n_e(r)$. Note that to deal with the light U(1) model case, we can perform some simple replacements: $\varepsilon_{ij}^f\rightarrow \delta^i_j \frac{g_{Z^{\prime}}^{2} Q_{i}^{\prime}Q_{f}}{\sqrt{2} G_{F}m_{Z^{\prime}}^{2}} $ for \UUniversal and \UBL, $\varepsilon _{ij}^{f}\rightarrow \delta _{j}^{i}\frac{\sqrt{2} \alpha _{\mathrm{EM}} g_{Z^{\prime }}^{2} (\delta _{i\mu } \varepsilon _{\tau \mu } (0)+\delta _{i\tau } \varepsilon _{\mu \tau } (0)) Q_{em}^{f}}{\pi G_{F} m_{Z^{\prime }}^{2}}$ for \ULTau, where $Q_{em}^{f}$ is the electric charge for the lepton or quark. The momentum transfer vanishes since the forward scattering does not change the momenta of particles. For light U(1) models, there are only new diagonal terms in $\boldsymbol{H_\text{mat}}$.
According to the electrical neutrality, the up- and down-quark relative abundances can be written in terms of the neutron relative abundance $Y_n$ according to $Y_u = 2 + Y_n, Y_d = 1 + 2 Y_n$. Neglecting metallicity elements in the sun, $Y_n$ can be computed by $n_n(r)\simeq\frac{X(^4\text{He})}{2X(^1\text{H})+X(^4\text{He})}$, where $X(^1\text{H})$ and $X(^4\text{He})$ are the mass fractions. All the standard solar model (SSM) related quantities that we utilized can be found in the BS05 SSM~\cite{Bahcall:2004pz}.}

\zbl{In the mass dominance limit, i.e.  $\Delta m_{31}^2\to\infty $, the three neutrino oscillation is reduced to the simple two neutrino oscillation. Using the propagation basis $|\boldsymbol{\nu}\rangle \rightarrow |\boldsymbol{\tilde  \nu}\rangle = \boldsymbol{U}(\theta_{13})^T \boldsymbol{U}(\theta_{23})^T|\boldsymbol{\nu}\rangle$,  the new $\boldsymbol{\tilde H_\text{mat}}$ can be viewed as the direct sum of a $2\times2$ matrix and a  $1\times1$ matrix, since $\tilde{\nu}_e-\tilde{\nu}_\tau$ and $\tilde{\nu}_\mu-\tilde{\nu}_\tau$ mixing are suppressed by $\Delta m_{31}^2$. Therefore, there exists only the $\tilde{\nu}_e-\tilde{\nu}_\mu$ oscillation. 
Let us focus on the Hamitonian of the $\tilde{\nu}_e-\tilde{\nu}_\mu$ oscillation:
$\boldsymbol{H}=\frac{1}{4E_\nu}  \begin{pmatrix}    -\Delta m_{21}^2 \cos2\theta_{12} + A     & \Delta m_{21}^2 \sin2\theta_{12} + B\\    \Delta m_{21}^2 \sin2\theta_{12} + B     & \Delta m_{21}^2 \cos2\theta_{12} - A  \end{pmatrix}\,,$
where $$A=4\sqrt{2}E_\nu G_F n_e(r)   \left[\frac{\cos^2\theta_{13}}{2} - Y_q(r)\varepsilon_D\right]\ ,\qquad  B=4\sqrt{2}E_\nu G_F n_e(r) Y_q(r)\varepsilon_N\,,$$with $$\begin{aligned}\varepsilon_D&=-\frac{c_{13}^2}{2}\epsilon_{ee}^q+\frac{\left[c_{13}^2-\left(s_{23}^2-s_{13}^2c_{23}^2\right)\right]}{2}\epsilon_{\mu\mu}^q+\frac{\left(s_{23}^2-c_{23}^2s_{13}^2\right)}{2}\epsilon_{\tau\tau}^q+s_{13}c_{13}s_{23}\epsilon_{e\mu}^q\\&+s_{13}c_{13}c_{23}\epsilon_{e\tau}^q- (1+s13^2)c_{23}s_{23}\epsilon_{\mu\tau}^q,\\\varepsilon_N&=-s_{13}c_{23}s_{23}\epsilon_{\mu\mu}^q+s_{13}c_{23}s_{23}\epsilon_{\tau\tau}^q+c_{13}c_{23}\epsilon_{e\mu}^q-c_{13}s_{23}\epsilon_{e\tau}^q+s_{13}\left(s_{23}^2-c_{23}^2\right)\epsilon_{\mu\tau}^q\,.\end{aligned}$$
Here, $\cos\theta_{ij}=c_{ij}$ and $\sin\theta_{ij}=s_{ij}$. We do not consider the CP-violating phase, while this effect is considered in Ref.~\cite{Gonzalez-Garcia:2013usa}~\footnote{We modify the expression of $\varepsilon_D$ in Ref.~\cite{AristizabalSierra:2017joc} by $s_{13}c_{13}c_{23}\epsilon_{e\mu}^q \rightarrow  s_{13}c_{13}c_{23}\epsilon_{e\tau}^q$ and $- c_{23}s_{23}\epsilon_{\mu\tau}^q \rightarrow - (1+s13^2)c_{23}s_{23}\epsilon_{\mu\tau}^q$, so that our derivation is consistent with that of~Ref.~\cite{Gonzalez-Garcia:2013usa}.}. The effective mixing angle in matter $\theta_M$ can be obtained by:
$$\cos2\theta_M(r) = \frac{\Delta m_{12}^2\cos2\theta_{12}-A}  {\sqrt{\left(\Delta m_{12}^2\cos2\theta_{12}-A\right)^2      +  \left(\Delta m_{12}^2\sin2\theta_{12}+B\right)^2}}\,.$$ The survival probability of the solar neutrino is given by~\cite{Kuo:1986sk,Parke:1986jy} $$P_{ee}=\cos^4\theta_{13}P_{eff}+\sin^4\theta_{13}\,,$$
where
$P_{eff}= ( 1+ \cos 2\theta_{\mathrm{M} }\cos 2\theta_{12}) /2$ is the effective probability relying on the propagation path and the matter effect. Other survival probabilities include $P_{e\mu}= ( 1- P_{ee}) \cos ^2\theta_{23}, $ and $P_{e\tau}= $ $( 1- P_{ee}) \sin ^2\theta_{23}$. }

\section{\zblt{Constraints on U(1) models}}
\label{appendix5}
\begin{figure}[htbp]
	\centering
	\includegraphics[width=0.7\linewidth]{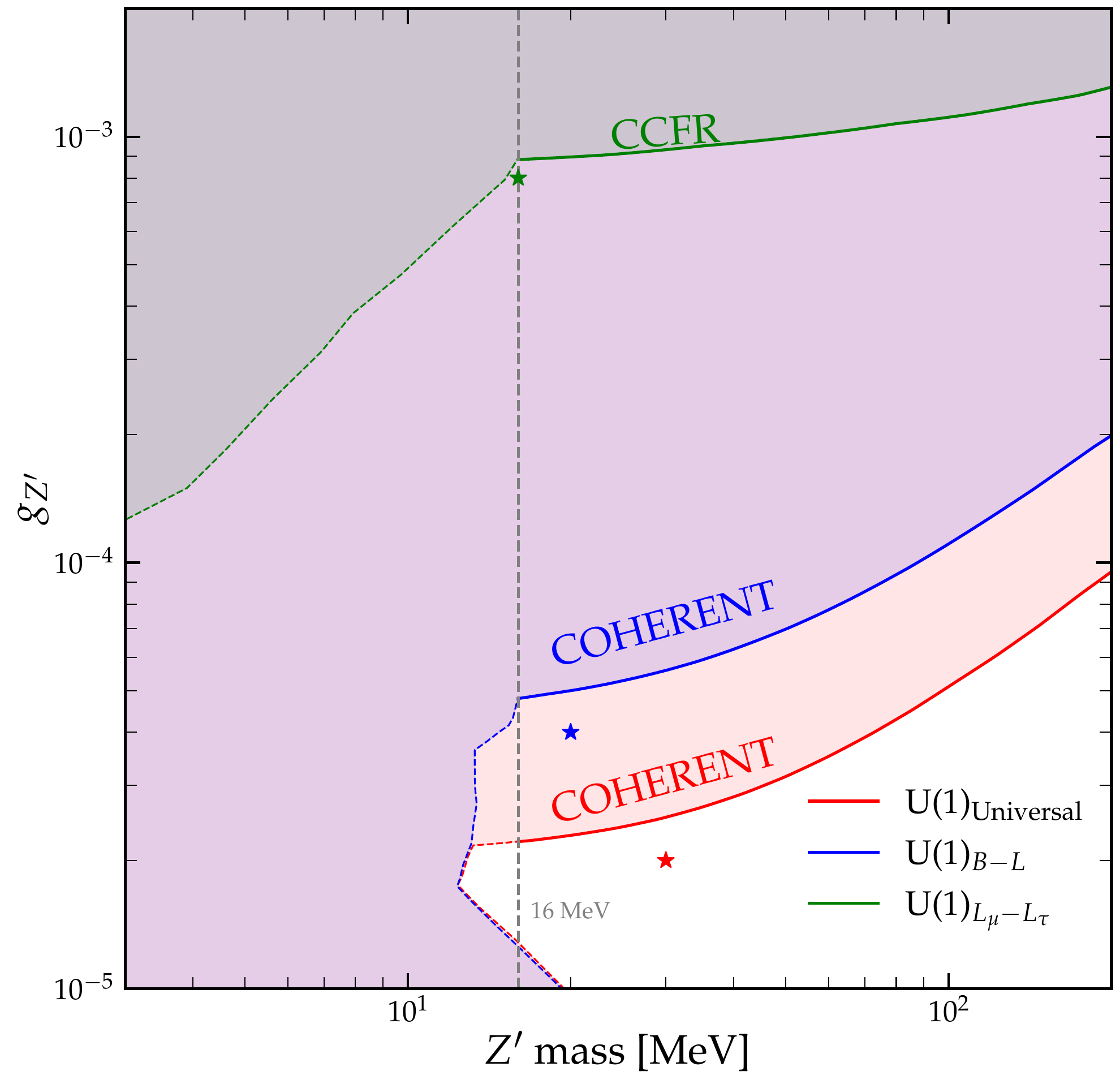}
	\caption{\label{fig:U1-constraints} \zblt{The $2\sigma$ excluded regions for U(1) models from Ref.~\cite{AtzoriCorona:2022moj}. Solid lines represent U(1) parameters adopted in Part~\ref{sec:E}. The three benchmark points are marked by stars.}}
\end{figure}
\zblt{For simplicity, we only show the constraints on U(1) models for $\rm 3~\mathrm{MeV}\lesssim m_{Z^\prime}\lesssim 200~\mathrm{MeV}$ in Fig.~\ref{fig:U1-constraints}. In our regions of interest, $\rm 16~\mathrm{MeV}\lesssim m_{Z^\prime}\lesssim 200~\mathrm{MeV}$, the constraints on \UUniversal and \UBL models are dominated by COHERENT datasets~\cite{AtzoriCorona:2022moj}, while the constraints on \ULTau model mainly come from neutrino trident production processes~\cite{CCFR:1991lpl,Altmannshofer:2014pba}. Below $m_{Z^\prime}\simeq 16~\mathrm{MeV}$, the strongest constraints on \UUniversal and \UBL models are provided by the combinations of beam-dump experiments, while the solar neutrino observatory together with cosmological observations give the strongest limits on \ULTau model. Above $m_{Z^\prime}\simeq 200~\mathrm{MeV}$, the strongest constraints on three U(1) models are derived from the combinations of collider experiments.}

\section{\zbl{Results with different thresholds}}
\label{appendix4}
\begin{figure}[htbp]
	\centering
	\includegraphics[width=0.9\linewidth]{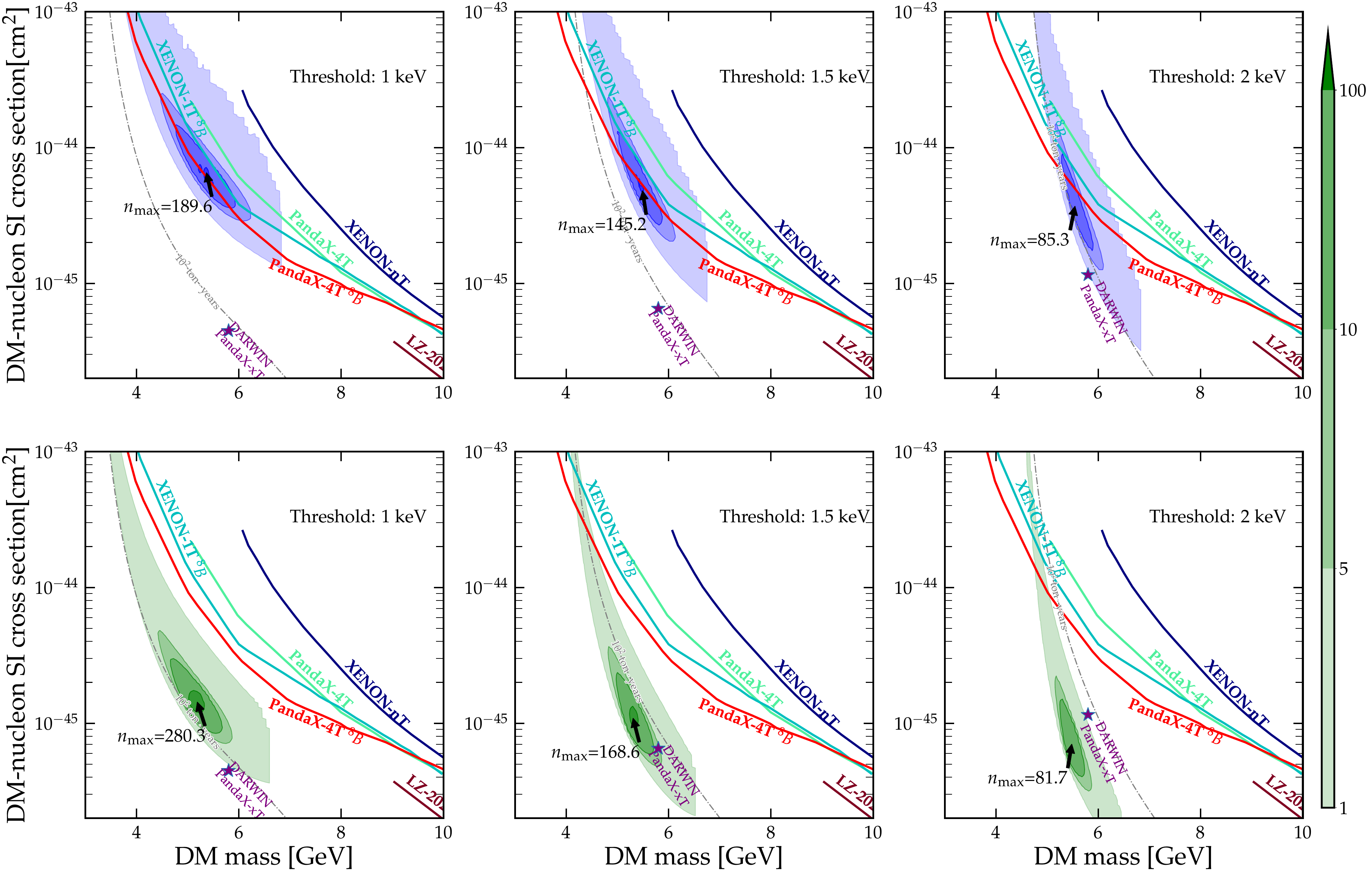}
	\caption{\label{fig:DMN_Threshold_2} \zbl{The distinction plot for DM-nucleon interaction in the $\rm DM~v.s.~\nu NP$ scenario.}}
\end{figure}

\begin{figure}[htbp]
	\centering
	\includegraphics[width=0.9\linewidth]{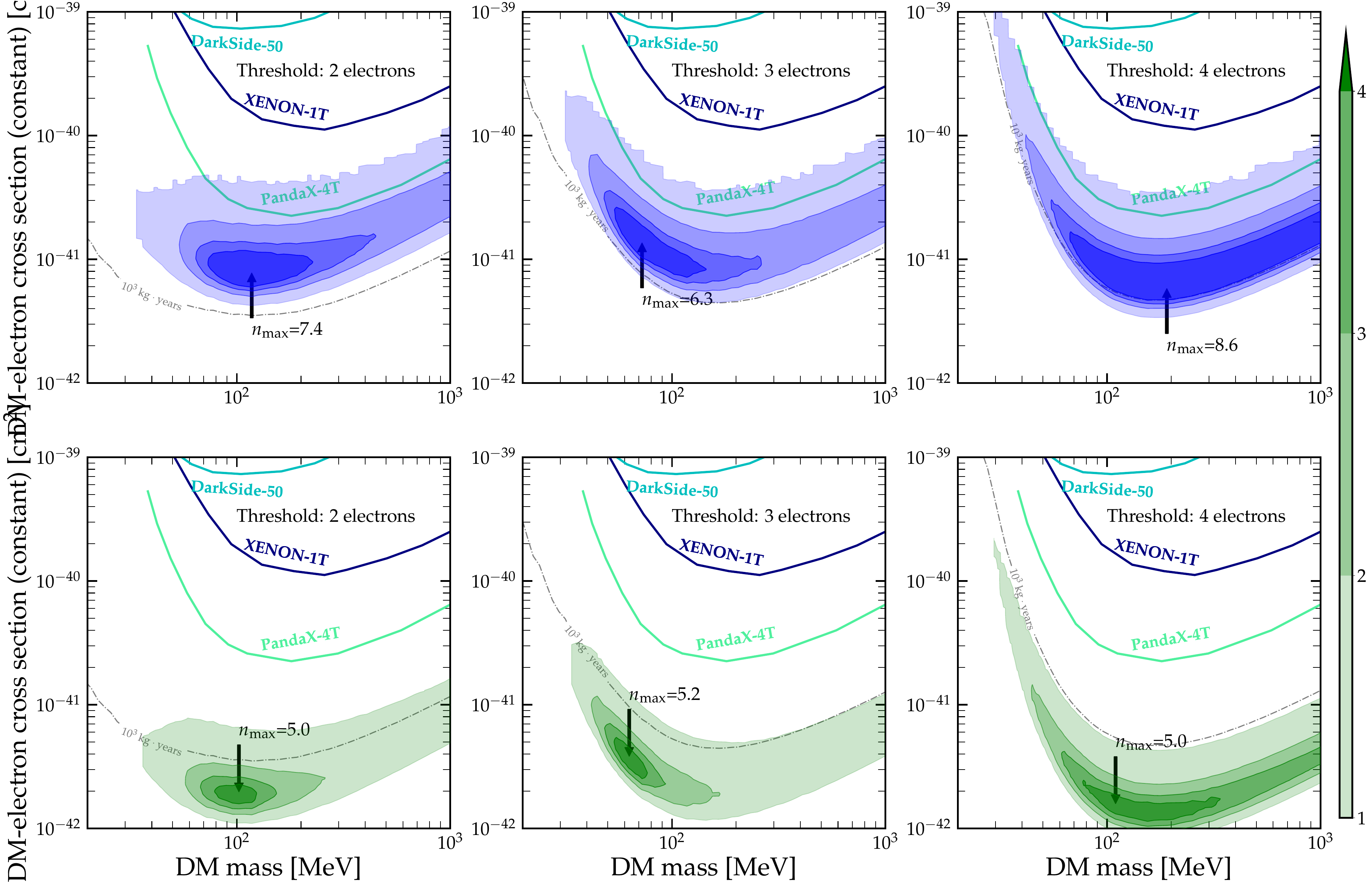}
	\caption{\label{fig:DMe_Threshold_2} \zbl{The distinction plot for DM-electron interaction in the $\rm DM~v.s.~\nu NP$ scenario.}}
\end{figure}

\zbl{As detailed in Part~\ref{sec:D}, the spectra of DM and neutrino are modified with different thresholds, leading to a varying discrepancy between spectra under two statistical hypotheses. As shown in Fig.~\ref{fig:DMe_Threshold_2}, the distinction regions for the DM-nucleon interaction become narrower and shallower as the energy threshold rises. Furthermore, a higher threshold imposes limitations on detecting weaker DM-nucleon interactions, particularly with a fixed exposure time.
Therefore, despite the advent of low-energy threshold detection techniques and substantial progress in statistical data analysis, experimental physicists must remain vigilant in accounting for the potential influence of $\nu$NP when interpreting their findings.}

\zbl{Fig.~\ref{fig:DMe_Threshold_2} illustrates that the trend observed for the \UBL model closely resembles that of the \UUniversal model, whereas the trend for the \ULTau model exhibits a contrasting behavior. For the \ULTau model, a threshold of three emitted electrons yields the more degenerate spectra obtained under two hypotheses, as evidenced by the variation in $n_\mathrm{max}$, compared to a higher threshold of four emitted electrons. Nevertheless, reducing the threshold from three to two emitted electrons results in an amplification in the discrepancy between spectra.
Hence, it is crucial to invest efforts in optimizing the detector threshold in DM experiments, as this will aid in quantifying the influence of $\nu$NP for the DM-electron interaction scenarios.}




\bibliographystyle{JHEP}
\bibliography{biblio.bib}


\end{document}